\documentclass[%
 reprint,%
 amssymb, amsmath,%
 aip,cha,%
]{revtex4-1}

\usepackage{docs}%
\usepackage{bm}%
\usepackage[colorlinks=true,linkcolor=blue]{hyperref}%
\expandafter\ifx\csname package@font\endcsname\relax\else
 \expandafter\expandafter
 \expandafter\usepackage
 \expandafter\expandafter
 \expandafter{\csname package@font\endcsname}%
\fi
\usepackage{graphicx}
\usepackage{amssymb}
\DeclareMathOperator{\sech}{sech}

\begin{document}

\title{Soliton Interaction with External Forcing within the Korteweg--de Vries Equation \\ \phantom{Free space}}%

\author{Andrei Ermakov}
\affiliation{School of Agricultural, Computational and Environmental Sciences,\\ University of Southern Queensland, QLD
4350, Australia}%
\author{Yury Stepanyants}%
\email{Yury.Stepanyants@usq.edu.au}
\affiliation{School of Agricultural, Computational and Environmental Sciences,\\ University of Southern Queensland, QLD
4350, Australia and \\ Department of Applied Mathematics, Nizhny Novgorod State Technical University, \\ Nizhny Novgorod, 603950, Russia \\ \phantom{Free space}}%

\date{September 2018}%
\revised{December 2018 \\ \phantom{Free space}}%

\begin{abstract}
We revise the solutions of the forced Korteweg--de Vries equation describing a resonant interaction of a solitary wave
with external pulse-type perturbations. In contrast to previous works where only the limiting cases of a very narrow forcing in comparison with the initial soliton or a very narrow soliton in comparison with the width of external perturbation were studied, we consider here an arbitrary relationship between the widths of soliton and external perturbation of a relatively small amplitude. In many particular cases, exact solutions of the forced Korteweg--de Vries equation can be obtained for the specific forcings of arbitrary amplitude. We use the earlier developed asymptotic method to derive an approximate set of equations up to the second-order on a small parameter characterising the amplitude of external force. The analysis of exact solutions of the derived equations is presented and illustrated graphically. It is shown that the theoretical outcomes obtained by asymptotic method are in a good agreement with the results of direct numerical modelling within the framework of forced Korteweg--de Vries equation.
\end{abstract}

\maketitle

\section{Introduction}
\label{Intro}

The forced Korteweg--de Vries (fKdV) equation is a canonical model
for the description of resonant excitation of weakly nonlinear
waves by moving perturbations. Such an equation was derived by
many authors for atmospheric internal waves over a local
topography \cite{Patoine-1982, Warn-1983},
surface and internal water waves generated by moving atmospheric
perturbations or in a flow over bottom obstacles \cite{Akylas-1984, Cole-1985, Grimshaw-1986, Grimshaw-1987, Melville-1987, Smyth-1987, Wu-1987, Lee-1989, Grimshaw-1991, Mitsudera-1991}, internal
waves in a rotating fluid with a current over an obstacle
\cite{Grimshaw-1990}. The number of publications on this topic is so
huge that it is impossible to mention all of them in this article.
In addition to the papers mentioned above we will only add a review
paper \cite{Grimshaw-2010} and relatively recent publication
\cite{GrimMal-2016} where a reader can find some more refe\-rences.

An effective method of asymptotic analysis of fKdV
equation, when the amplitude of external force acting on a KdV
soliton is relatively small, was developed in the series of papers by Grimshaw and Pelinovsky with co-authors
\cite{GPT1994, GPS1996, GPB1997, Pelin2000, GrimPel2002}. Two limiting cases were analysed in
those papers: (i) when the width of external force is very small
in comparison with the width of a soliton and can be approximated
by the Dirac delta-function and (ii) when a soliton width is very
small in comparison with the width of external perturbation. Similar approach was used in Ref. \cite{Malomed1988} where the forcing term was approximated by the derivative of Dirac delta-function.

In the meantime, in the natural conditions a relationship between
the widths of solitary wave and external forcing can be arbitrary, therefore it is of
interest to generalise the analysis of those authors and consider
a resonance between the solitary waves and external forces
of arbitrary width. For such arrangements we have found few physically interesting regimes, which were missed in the previous studies. In
addition to that, we show that for some special external forces
exact solutions of fKdV equation can be obtained even when the
amplitude of external force is not small. We compare our solutions derived by means of asymptotic method with the results of direct numerical modelling within the framework of fKdV equation and show that there is a good agreement between two approaches. In the meantime, the numerical simulation demonstrates that there are some effects, which are not caught by the asymptotic theory.

Below we briefly describe the basic model and asymptotic method developed in the papers \cite{GPT1994, GPS1996, GPB1997, Pelin2000, GrimPel2002} for the analysis of soliton interaction with external forcing, and then we apply the basic set of approximate equations to the particular cases of stationary and periodic forcing. In Sect. \ref{NumRes} we present the results of numerical modelling and comparison of theoretical outcomes with the numerical data. In the Conclusion, we discuss the results obtained in this paper.

\section{The basic model equation and perturbation scheme}
\label{Sect-2}

In this paper we follow the asymptotic method deve\-loped in the
aforementioned papers \cite{GPT1994, GPS1996, GPB1997, Pelin2000,
GrimPel2002} and apply it to the fKdV equation in the form:
\begin{equation} %
\label{Eq01}%
\frac{\partial u}{\partial t} +c\frac{\partial u}{\partial x}
+\alpha u \frac{\partial u}{\partial x} +\beta \frac{\partial ^{3}
u}{\partial x^{3} } = \varepsilon\frac{\partial f}{\partial x},
\end{equation}
where $c$, $\alpha$ and $\beta$ are constant coefficients, and
$f(x, t)$ describes the external perturbation of amplitude $\varepsilon$
moving with the constant speed $V$.

Introducing new variables $\hat{x} = x - Vt$, $\hat{t} = t$, we can
transform Eq.~(\ref{Eq01}) to the following form (the symbol $\hat{}$
is further omitted):
\begin{equation} %
\label{Eq02}%
\frac{\partial u}{\partial t} +(c-V)\frac{\partial u}{\partial x}
+\alpha _{} u\frac{\partial u}{\partial x} +\beta \frac{\partial
^{3} u}{\partial x^{3} } = \varepsilon\frac{\partial f}{\partial x}.
\end{equation}

This form corresponds to the moving coordinate frame where the
external force is stationary and depends only on spatial
coordinate $x$.

In the absence of external force, i.e., when $f(x, t) \equiv 0$, Eq.~(\ref{Eq02}) reduces to the well-known KdV equation which has
stationary solutions in the form of periodic and solitary waves.
We study here the dynamics of a solitary wave under the action of
an external force of small amplitude $\varepsilon \ll 1$ assuming that in
the zero approximation (when $\varepsilon = V = 0$) the solution is
\begin{equation}
\label{Eq03}%
u_{0} = A_0 \sech^{2}(\gamma_0 \Phi),
\end{equation}
where the inverse half-width of a soliton $\gamma_0 = \sqrt{\alpha
A_0/12\beta\vphantom{a^2}}$ and its speed $\upsilon_0 = c + \alpha A_0/3$ depend on the amplitude $A_0$, $\Phi = x - x_0 - \upsilon_0t$ is the total phase of the soliton, and $x_0$ is
an arbitrary constant determining the initial soliton position at $t = 0$.

In the presence of external force of a small amplitude the
solitary wave solution (\ref{Eq03}) is no longer valid, but one
can assume that under the action of external perturbation it will
gradually vary so that its amplitude and other parameters can be considered as functions of ``slow time'' $T = \varepsilon t$, so that
\begin{eqnarray}
\upsilon(T) &=& c - V + \frac{\alpha A(T)}{3}, \label{Eq04a} \\%
\Psi(T) &=& x_{0} + \frac{1}{\varepsilon}
\int\limits_{0}^{T}\upsilon(\tau)\,d\tau. \label{Eq04b}%
\end{eqnarray}

Now we have to define functions $A(T)$ and $\upsilon(T)$. This can
be done by means of the asymptotic method deve\-loped, in
particular, in Refs. \cite{GrimMitsud1993, GPT1994}. Following these
papers, we seek for a solution of the perturbed KdV equation
(\ref{Eq02}) in the form of the expansion series:
\begin{equation} %
\label{Eq05}%
\begin{array}{l}
{u=u_{0} + \varepsilon u_{1} + \varepsilon^{2} u_{2} + \ldots} \\
{}\\
{\upsilon = \upsilon_{0} + \varepsilon \upsilon_{1} + \varepsilon^{2}
\upsilon_{2} + \ldots}
\end{array}
\end{equation}

In the leading order of perturbation method (in the zero
approximation), when $\varepsilon = 0$, we obtain the solitary wave
solution (\ref{Eq03}) for $u_{0}$ and $\upsilon_{0}$. In the next
approximation we obtain the same solution, but with slowly varying
parameters in time. The dependence of soliton amplitude $A$ on $T$ can be found from the energy balance equation \cite{GPT1994}, which follows from Eq. (\ref{Eq02}) after multiplication by $u(x, t)$ and integration over $x$:
\begin{equation} %
\label{Eq06} %
\frac{d}{dT} \int\limits_{-\infty }^{\infty }\frac{u^2(\Phi
)}{2} d\Phi = \int\limits_{-\infty }^{\infty }u(\Phi
)\frac{df(\Phi)}{d\Phi } d\Phi.
\end{equation}

Substituting here solution (\ref{Eq03}), we obtain the equations for $A(T)$:
\begin{equation} %
\label{Eq07a}%
\frac{dA}{dT} = \gamma\int\limits_{-\infty }^{\infty
}\sech^2{(\gamma\Phi)}\frac{df(\Phi + \Psi)}{d\Phi } d\Phi,
\end{equation}

The second equation for $\Psi(T)$ in this approximation represents just a kinematic condition: the time derivative of soliton phase is equal to the instant soliton speed in the moving coordinate frame:
\begin{equation} %
\label{Eq07b}%
\frac{d\Psi }{dT} = \Delta V + \frac{\alpha A(T)}{3},
\end{equation}\\
where $\Delta V = c - V$.

In the second order of asymptotic theory, a correction to the wave speed $\upsilon_1$ (see Eq. (\ref{Eq05})) should be taken into account. Leaving aside the derivation of the corrected equation (\ref{Eq07b}) (the details
can be found in \cite{GPT1994}), we present here the final equation:
\begin{widetext}
\begin{equation} %
\label{Eq08} %
\frac{d\Psi }{dT} = \Delta V + \frac{\alpha A(T)}{3} +
\frac{\varepsilon\alpha}{24\beta \gamma^{2} }
\int\limits_{-\infty}^{\infty}\left[\tanh{\gamma\Phi} + \left(\gamma \Phi - 1\right)\sech^2{\gamma\Phi}\right]\frac{\partial
f(\Phi + \Psi)}{\partial \Phi} d\Phi.
\end{equation}
\end{widetext}
	
Thus, the set of equations in the first approximation consists of
Eqs.~(\ref{Eq07a}) and (\ref{Eq07b}), whereas in the second
approximation it consists of Eqs.~(\ref{Eq07a}) and (\ref{Eq08}). However, as has been shown in Ref. \cite{GPT1994}, the last term in Eq.~(\ref{Eq08}) containing small parameter $\varepsilon$ dramatically changes the behaviour of the system and makes the result realistic, whereas Eq.~(\ref{Eq07b}) provides just a rough approximation to the real solution valid at fairly small time interval in the vicinity of a forcing. This difference between the solutions in the first and second approximations will be illustrated in the next Section, and then we will analyse only solutions corresponding to the second approximation described by Eqs. (\ref{Eq07a}) and (\ref{Eq08}) for different kinds of external force $f(x)$.

\section{The KdV-type forcing}
\label{Case1}

Let us consider first the case when
\begin{equation} %
\label{Eq09} %
f(x) = \sech^{2}{\frac{x}{\Delta_f}}, \quad V = c + \frac{4\beta}{\Delta_f^2} - \frac{\varepsilon \alpha\Delta_f^2}{12\beta},
\end{equation}
where $\Delta_f$ is a free parameter characterising the half-width of external
force.

With this function $f(x)$ one can find an exact solution of Eq.~(\ref{Eq02}) in the form of a KdV soliton (\ref{Eq03})
synchronously moving with the external force, $\upsilon_s = V$, and having the amplitude $A_s = 12\beta/\alpha\Delta_f^2$ and half-width $\gamma_s^{-1} = \Delta_f$. This solution represents a particular case of a family of exact solutions to
the class of forced generalised KdV equations constructed in
Ref.~\cite{LZM2001}. Note that here the parameters $\varepsilon$ and $\Delta_f$ are
arbitrary, and the amplitude $A_s$ of a soliton is determined only by the width of external force $\Delta_f$, whereas the soliton speed $V$ is determined both by the width $\Delta_f$ and amplitude $\varepsilon$ of external force.

\begin{figure}[t!]
	\centerline{\includegraphics[width=1.7in]{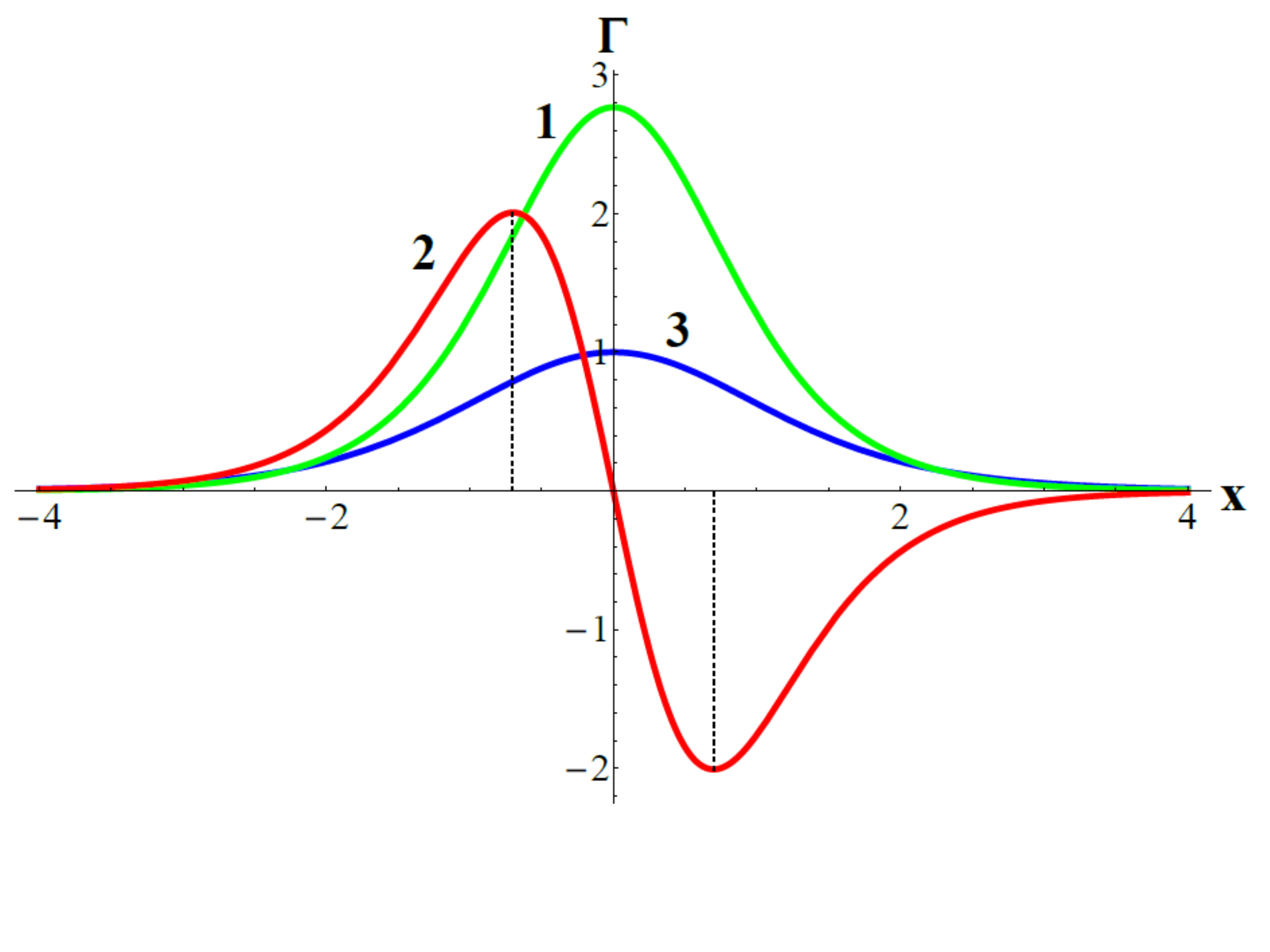} \hspace{-0.3cm} \includegraphics[width=1.7in]{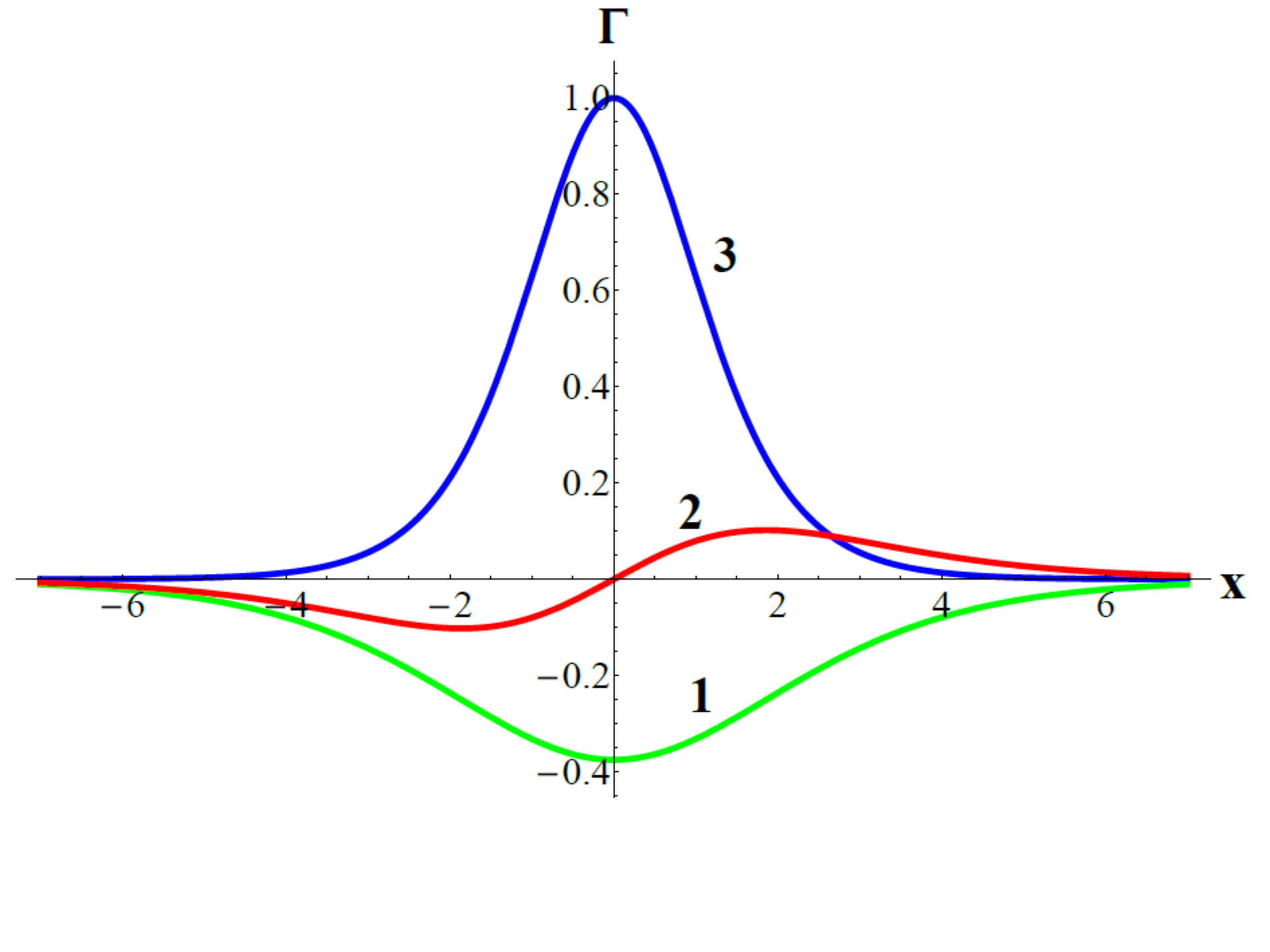}} %
	\vspace*{-0.5cm}
	{\hspace*{0.5cm} a) \hspace{3.5cm} b) \hspace{0.5cm}}%
	\vspace*{-0.1cm}
	\caption{The shape and polarity of forcing function $f(x)$ (green lines 1) as per Eqs. (\ref{Eq09}) and (\ref{Eq121}) for $K = 0.75$ (a) and $K = 2$  (b), red lines 2 represent the derivatives $f'_x(x)$, and blue lines 3 show the initial KdV solitons of unit amplitudes.} %
	\label{ErmStep_Fig_1} %
	\vspace{0.3cm}%
	\centerline{\includegraphics[width=1.6in]{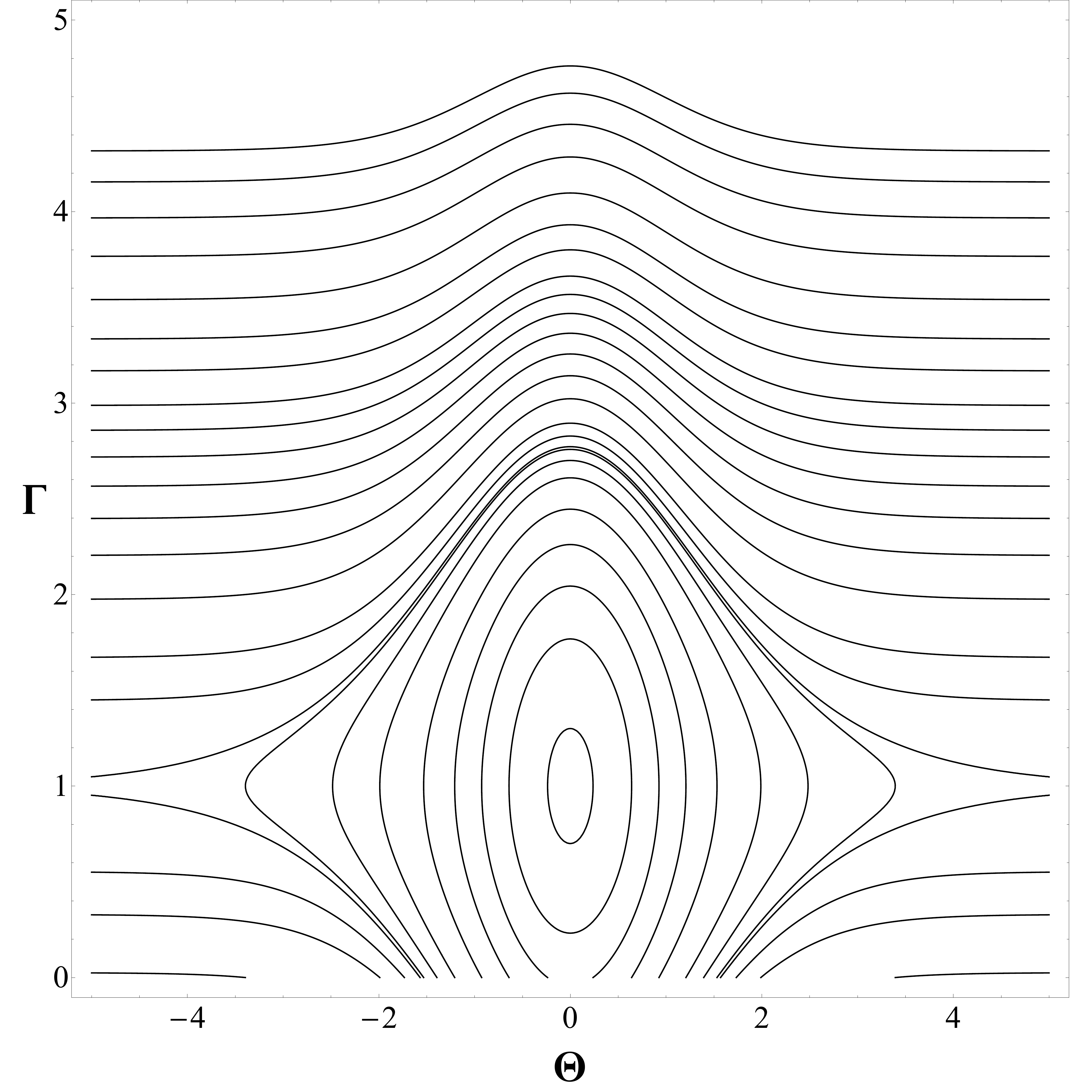}\hspace{1mm} \includegraphics[width=1.6in]{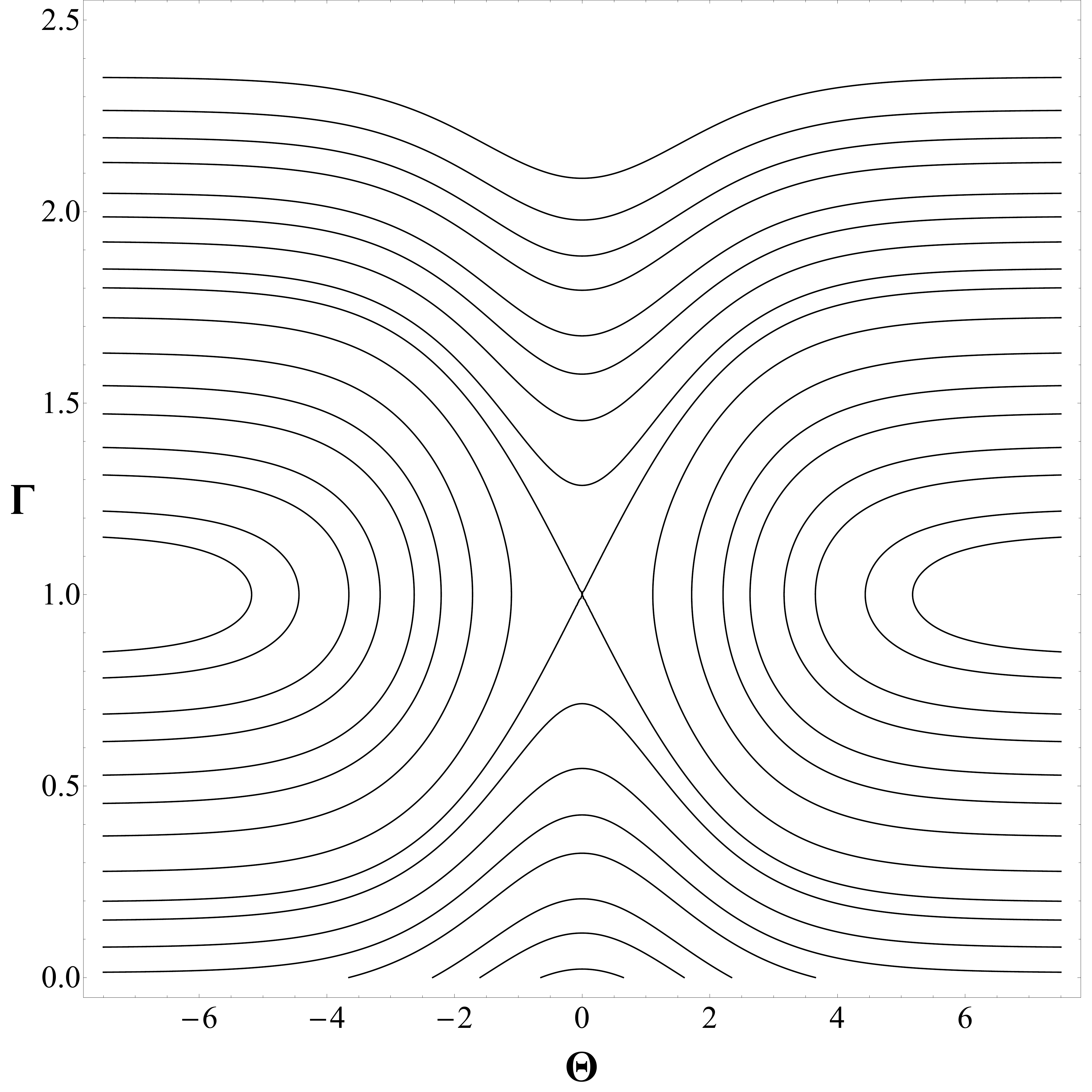}} %
	{\hspace{1.8cm} a) \hspace{3.9cm} b)}
	\vspace*{-0.1cm}
	\caption{The phase portraits of the dynamical system (\ref{Eq11})--(\ref{Eq12}) as per Eq.~(\ref{Eq14}) in the first approximation on the parameter $\varepsilon$ for $K = 0.75$ (a) and $K = 2$ (b).} %
	\label{ErmStep_Fig_2} %
	\vspace{0.3cm}%
	\centerline{\includegraphics[width=1.6in]{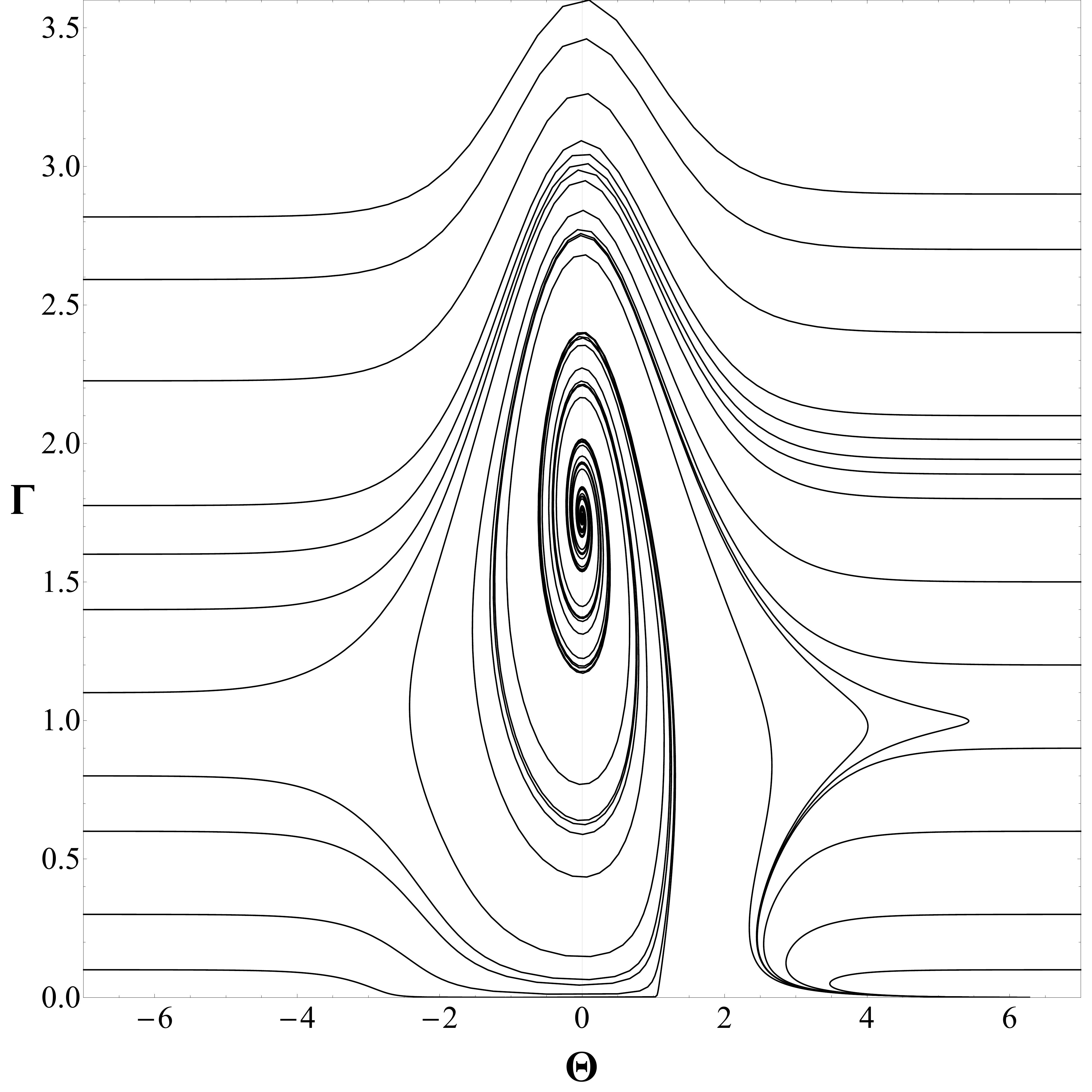}\hspace{1mm} \includegraphics[width=1.6in]{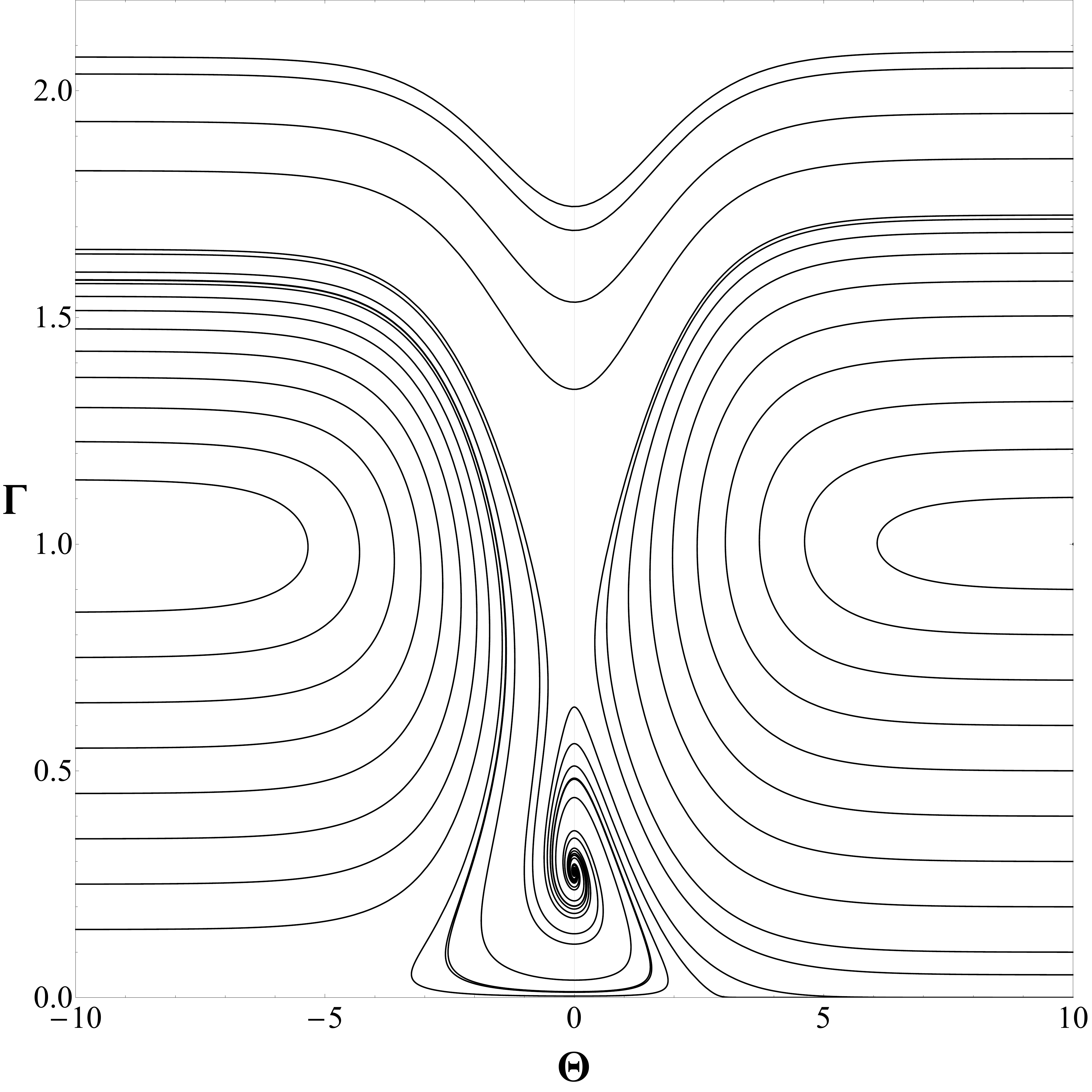}} %
	{\hspace{1.8cm} a) \hspace{3.9cm} b)}
	\vspace*{-0.1cm}
	\caption{The phase portraits of the dynamical system (\ref{Eq11}), (\ref{Eq19}) in the second approximation on the parameter $\varepsilon$ for $K = 0.75$ (a) and $K = 2$ (b).} %
	\label{ErmStep_Fig_3}
\end{figure}

Let us assume now that the parameter $\varepsilon$ is small, and we
have the initial condition for Eq.~(\ref{Eq02}) in the form of KdV
soliton shifted from the centre of forcing and moving with its own velocity $\upsilon_{0}$ with the initial amplitude $A_0 \ne A_s$. By substitution of function $f(x)$ from Eq.~(\ref{Eq09}) in
Eq.~(\ref{Eq07a}), we obtain for the parameter $\gamma(T)$ the
following equation:
\begin{equation} %
	\label{Eq11} %
	\frac{d\gamma}{dT} = -\frac{2 \varepsilon \alpha}{3 \beta}e^{2\theta} \int
	\limits_{0}^{\infty} \frac{q^{K}}{\left(e^{2\theta } +
		q^{K}\right)^{2}} \frac{q - 1}{\left(q + 1\right)^{3}}dq,
\end{equation}
where $q =
\exp{\left(2\Phi/\Delta_f\right)}$, $\theta = \gamma\Psi$, and $K = \gamma_0\Delta_f$ is the ratio of half-widths of
external force and initial soliton. The parameter $K$ can be also presented in terms of the half-distance $D_f$ between the extrema of forcing function $f(x)$: $K = 2 \gamma_0 D_f/\ln{(2 + \sqrt{3})}$ (see the distance between maximum and minimum of $f'_x$ in Fig. \ref{ErmStep_Fig_1}).

Equation (\ref{Eq07b}) of the first approximation in terms of $\theta = \gamma\Psi$ reads (cf. \cite{GPT1994}):
\begin{equation} %
	\label{Eq12} %
	\frac{d\theta}{dT} = \Delta V\gamma + 4\beta \gamma ^{3}.
\end{equation}

According to the asymptotic theory, soliton velocity should be close to the forcing velocity. If we assume that at the initial instant of time they are equal, $\upsilon_{0} = V$, then we obtain that the forcing amplitude $\varepsilon$ is linked with the initial soliton amplitude $A_0$ through the formula:
\begin{equation} %
	\label{Eq121} %
	\varepsilon = \frac{\alpha A_0^2\left(1 - K^2\right)}{3K^4}.
\end{equation}
This formula shows that the polarity of forcing depends on the sign of its amplitude $\varepsilon$ and is determined by the parameter $K$: it is positive, if $K < 1$, and negative other\-wise.

Dividing Eq.~(\ref{Eq11}) by Eq.~(\ref{Eq12}), we obtain:
\begin{equation} %
	\label{Eq13} %
	\frac{d\gamma}{d\theta} =
	-\frac{2\varepsilon\alpha e^{2\theta}}{3 \beta\gamma \left(\Delta V + 4\beta \gamma ^2\right)} \int \limits_{0}^{\infty}
	\frac{q^{K}}{\left(e^{2\theta } + q^{K}\right)^{2}} \frac{q -
		1}{\left(q + 1\right)^{3}}dq.
\end{equation}
This is the first-order separable equation whose general solution can be presented in the form:
\begin{widetext}
	\begin{equation} %
		\label{Eq14} %
		\Gamma^{2} + 2\Gamma = 32\frac{K^{2} - 1}{K^{4}} \int\int\limits_{0}^{\infty } \left[\frac{q^{K}}{\left(e^{2\theta} +
			q^{K}\right)^{2}}\frac{q - 1}{\left(q + 1\right)^{3}}dq\right]e^{2\theta}d\theta + C,
	\end{equation}
\end{widetext}
where $\Gamma = A/A_{0}$ is the dimensionless amplitude of a solitary wave, and $C$ is a constant of integration.

The integrals in the right-hand side of Eq.~(\ref{Eq14}) can be
evaluated analytically; however we do not present here the results of integration as they are very cumbersome. After evaluation of the integrals in Eq. (\ref{Eq14}), the phase portrait of the dynamical system (\ref{Eq11})--(\ref{Eq12}) in terms of the dependence $\Gamma(\theta)$ can be plotted for any value of the parameter $K$.

In the case when the width of initial solitary wave is the same as
the width of external force, i.e., $K = 1$, we obtain $\Gamma = 1$
and $C = 3$.

When $K$ varies in the range $0 < K < 1$, then the forcing is positive, $\varepsilon f(x)>0$
(see Fig.~\ref{ErmStep_Fig_1}a), and the right-hand side of
Eq.~(\ref{Eq14}) is positive too; then the equilibrium state with
$\Gamma = 1$ and $\theta = 0$ is of the centre-type in the phase
plane. Therefore, if soliton parameters are such that it is slightly shifted from the equilibrium position, then it will oscillate around this position as shown in the phase plane of Fig.~ \ref{ErmStep_Fig_2}a). This formally
corresponds to the trapping regime when a solitary wave is trapped in the neighbourhood of centre of external force.

If the amplitude and speed of initial soliton are big enough, then
the soliton simply passes through the external perturbation and
moves away. Such a regime of motion corresponds to the transient trajectories shown in the phase
plane of Fig.~ \ref{ErmStep_Fig_2}a) above the separatrix (the line dividing trapped and transient trajectories).

There are also trajectories in the lower part of the phase plane which
either bury into the horizontal axes with $\Gamma = 0$, or originate
from this axis. Such trajectories correspond to decay of solitons of certain amplitudes or birth of
solitons from small perturbations, which however appear for a while, but then decay. Some of these trajectory types, which appear within the separatrix, corres\-pond to the ``virtual solitons'' (see unclosed trajectories within the separatrix in Fig.~ \ref{ErmStep_Fig_2}a). The ``virtual solitons'' are generated in the neighbourhood of forcing maximum, then increase, but after a while completely disappear.

When $K > 1$, then the forcing is negative, $\varepsilon f(x) < 0$ (see
Fig.~\ref{ErmStep_Fig_1}b), the right-hand side of
Eq.~(\ref{Eq14}) is negative too, and the equilibrium state with
$\Gamma = 1$ and $\theta = 0$ is of the saddle-type, as shown in
Fig.~ \ref{ErmStep_Fig_2}b). In this case there are repulsive
regimes, where solitary waves approach the forcing either from the left or from the right and
bounce back. There are also the transient regimes above and below separatrices, where solitons of big
or small amplitudes simply pass through the forcing. There are regimes corresponding to the ``virtual solitons'', which arise for a while from small perturbations and then disappear (see the trajectories originating at the line $\Gamma = 0$ in Fig.~ \ref{ErmStep_Fig_2}b).

In this approximation our results are qualitatively si\-mi\-lar to the results obtained in Ref.~\cite{GPT1994}, but in contrast to that paper, as well as subsequent papers \cite{GPS1996,GPB1997,Pelin2000,GrimPel2002}, we do not use here the approximation of soliton or forcing by the Dirac delta-function. In the limiting cases, when the width of one of these entities becomes very small, our results completely reduce to those derived in Ref.~\cite{GPT1994}.

As was already noted in Ref.~\cite{GPT1994}, asymptotic equations of the first approximation actually do not provide physically realistic description of soliton dynamics. Only in the second approximation the dynamical system for $\Gamma$ and $\theta$ reflects a realistic description. In this approximation, Eq.~(\ref{Eq11}) remains the same, and Eq.~(\ref{Eq12}) should be replaced by a more complex equation, which follows from Eq.~(\ref{Eq08}) and in terms of function $\theta$ reads:
\begin{widetext}
	\begin{equation} %
		\label{Eq19} %
		{\frac{d\theta }{dT} =\Delta V\gamma + 4\beta \gamma ^{3} - \frac{\Delta V^{2}}{\beta \gamma } \frac{K^{2} - 1}{K^{4}} \int\limits_{0}^{\infty} \frac{e^{2\theta}\left(1 + 2\theta - K\ln{q}\right) - q^{K}}{\left(e^{2\theta} + q^{K} \right)^{2}} \frac{q - 1}{\left(q + 1\right)^{3}}q^{K}dq}.
	\end{equation}
\end{widetext}
The integral on the right-hand side of this equation can be calculated analytically, but the result is very cumbersome.

Combining Eq.~(\ref{Eq19}) with Eq.~(\ref{Eq11}), one can plot the improved phase portrait of the dynamical system; it is shown in Fig. \ref{ErmStep_Fig_3}. It is evident that the phase portrait in the second approximation dramatically differs from the phase portrait of the first approximation. First of all, the equilibrium state of the centre-type in Fig. \ref{ErmStep_Fig_2}a) maps into the unstable focus, alias spiral (see Fig. \ref{ErmStep_Fig_3}a); this has been noticed already in Ref. \cite{GPT1994}. Secondly, the equilibrium amplitude $\Gamma$ in the second approximation is greater than in the first approximation. Thirdly, on the transient trajectories of Fig. \ref{ErmStep_Fig_3}a) soliton amplitudes do not return back to their initial values (cf. asymptotics of transient trajectories above the focus, when $\theta \to \pm \infty$). There are some other important features which were missed in Ref. \cite{GPT1994} because of additional approximation of soliton or forcing by the Dirac delta-function. In particular, when $K < 1$, there is a repulsive regime clearly visible in the right lower corner of Fig. \ref{ErmStep_Fig_3}a). 

Similarly, there are differences in the phase portraits of first and second approximation when $K > 1$. In particular, a new equilibrium state of a stable focus appears below the saddle (which is not visible in Fig. \ref{ErmStep_Fig_3}b) due to rarefaction of trajectories, but clearly implied as a separator between the transient and captured trajectories). Note that in Ref. \cite{GPT1994} the focus was mistakenly identified with the centre-type equilibrium state. This equilibrium state corresponds to the small-amplitude soliton trapped by the negative forcing shown in \ref{ErmStep_Fig_1}b). Meanwhile, it is clear from the physical point of view and confirmed through the analysis of dynamical system in the second approximation that a positive forcing, such as shown in Fig. \ref{ErmStep_Fig_1}a), cannot trap and confine a soliton.

\section{The KdVB-type forcing}
\label{Case2}

In this section we consider Eq. (\ref{Eq02}) with the different and non-symmetric forcing function of the form:
\begin{equation} %
\label{Eq16}%
f\left(x\right) = \left(\pm 1 - \tanh{\frac{x}{\Delta_f}}\right)\sech^2{\frac{x}{\Delta_f}}.
\end{equation}

Equation (\ref{Eq02}) with this forcing function can be derived from the Korteweg--de Vries--Burgers (KdVB) equation and has the exact solution for any parameter $\Delta_{f}$ in the form of a shock wave \cite{W1996}:
\begin{equation} %
\label{ErmStep_Eq_18}%
u\left(x\right) = \varepsilon \Delta_f \left(1 \pm \tanh{\frac{x}{\Delta_f}} + \frac{1}{2}\sech^{2}{\frac{x}{\Delta_f}} \right),
\end{equation}
whereas the forcing amplitude $\varepsilon$ and speed $V$ are determined by the forcing width $\Delta_{f}$: 
$$
\varepsilon = \frac{24 \beta}{\alpha \Delta_f^3}; \quad V = c + \frac{24 \beta^2}{\Delta_f^2}.
$$

\begin{figure}[b!]
	\centerline{%
		\includegraphics[width=1.7in]{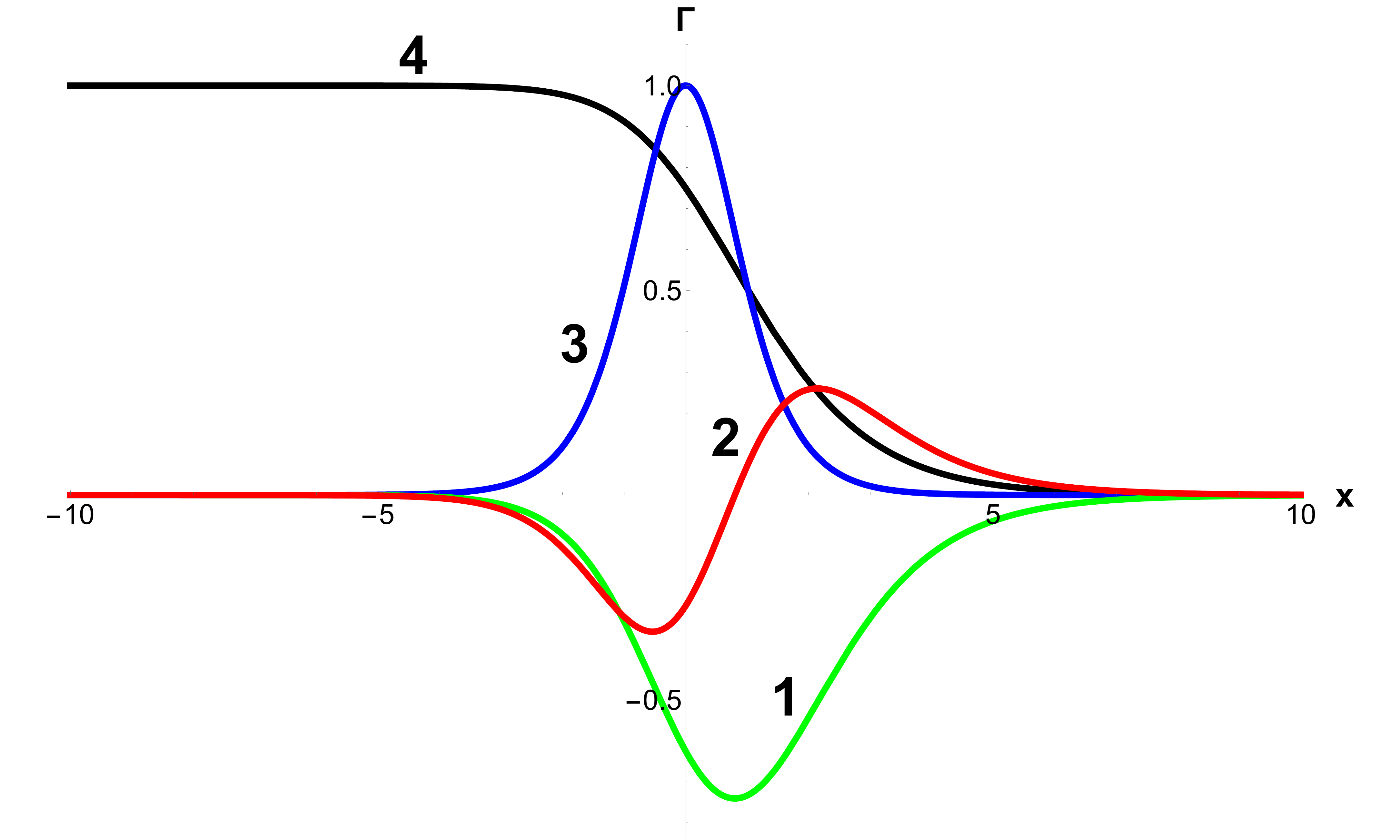}
		\includegraphics[width=1.7in]{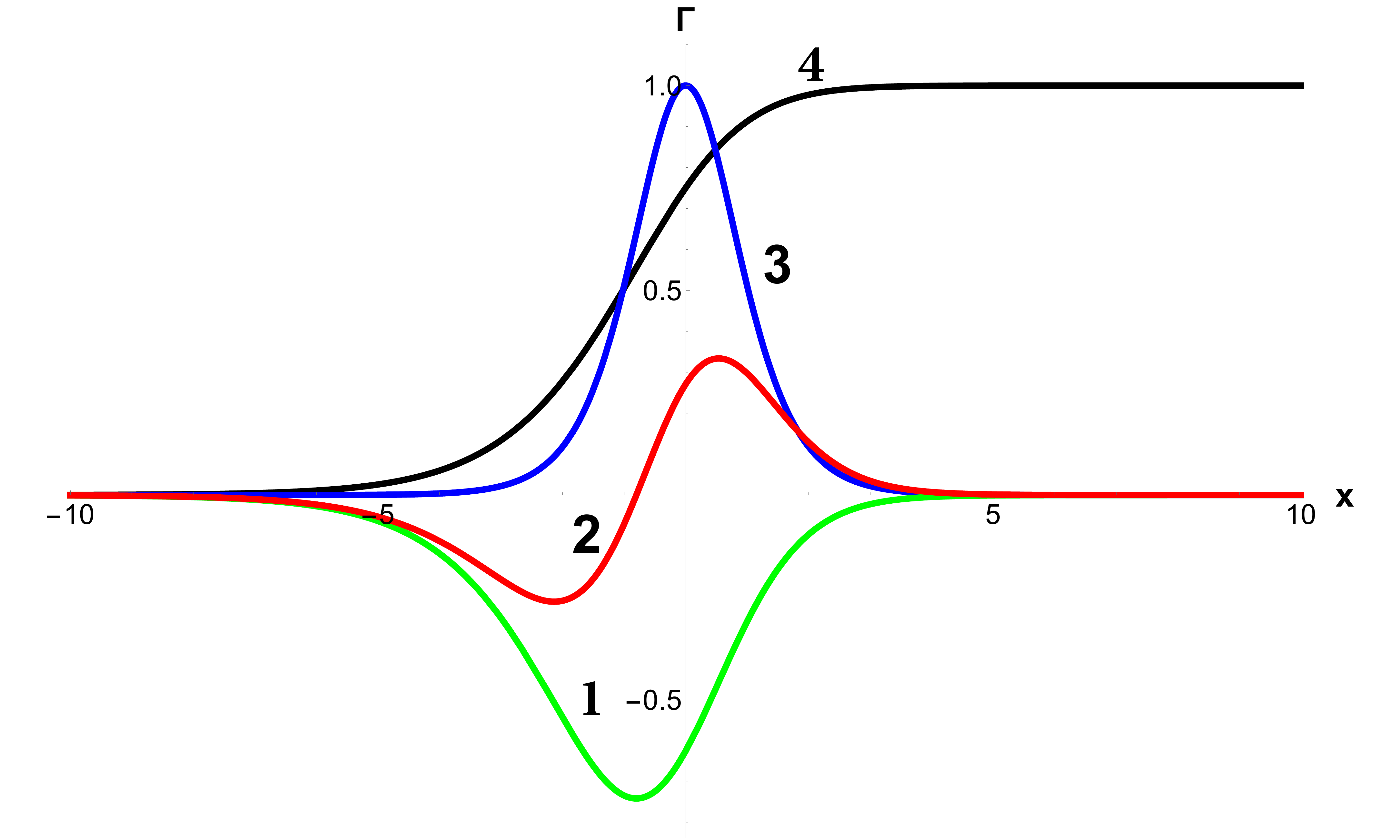}}
	{a) \hspace{4.1cm} b)}
	\caption{The forcing function with $K = 2$. Frame (a) pertains to the upper sign in Eq. (\ref{Eq16}), and frame (b) -- to the lower sign. Green lines (1) illustrate forcing functions $f(x)$, red lines (2) -- its derivatives $f'(x)$, blue lines 3 represent the KdV solitons at the initial instant of time, and black lines 4 represent the exact solutions of KdVB equation (\ref{ErmStep_Eq_18}) in the forms of shock wave (in frame a) and ``anti-shock wave'' (in frame b).} %
	\label{ErmStep_Fig_4}%
\end{figure}

The forcing function (\ref{Eq16}) and its derivative $f'_x$ are shown in Fig. \ref{ErmStep_Fig_4}. In the same figure one can see the exact solutions (\ref{ErmStep_Eq_18}) for the shock wave (black line in frame a) and ``anti-shock wave'' (black line in frame b). As follows from the exact solutions, a localised external force can produce a non-localised perturbation for $u(x)$ in the fKdV equation (\ref{Eq01}). Two different forcing functions correspon\-ding to the upper and lower signs in Eq.~(\ref{Eq16}) are mirror symmetric with respect to the vertical axis, therefore we illustrate below the solutions generated by only one of them shown in Fig. \ref{ErmStep_Fig_4}a), but for the sake of generality, below we present solutions for both signs in Eq. (\ref{Eq16}). Note that the forcing function (\ref{Eq16}) of any sign always represents only a negative potential shifted from the centre either to the right or to the left (see green lines 1 in the figure).

If the initial perturbation is chosen in the form of a KdV soliton (\ref{Eq03}) and the amplitude of external force is small, $\varepsilon \ll 1$, then we can apply again the asymptotic theory presented in Sect. \ref{Sect-2} to describe the evolution of a soliton under the influence of external force (\ref{Eq16}). In this case Eq. (\ref{Eq07a}) after substitution soliton solution and the forcing function (\ref{Eq16}) reduces to the following equation:

\begin{widetext}%
	\begin{equation} %
	\label{ErmStep_Eq_19}%
	\frac{d}{dT} \left(\frac{2A^{2} }{3\gamma } \right) = \pm \frac{10 \beta \varepsilon }{\Delta _{f}^2 }%
	\!\!\!\int\limits_{-\infty }^{\infty } \!\!A\sech^{2} (\gamma\Phi )\sech^{4} \left(\frac{\Phi +\Psi }{\Delta _{f} } \right) \left[2 - e^{\pm \frac{2(\Phi +\Psi )}{\Delta_{f}}}\right]d\Phi.
	\end{equation}
\end{widetext}

Introducing the parameters $q = e^{2\Phi /\Delta _{f}} $ and $K = 2 \gamma_0 D_f/\ln{[(7 + \sqrt{33})/4]}$, where $D_f$ as above, is the half-distance between the extrema of forcing derivative $f'_x$ (see Fig. \ref{ErmStep_Fig_4}) and skipping Eq. (\ref{Eq07b}) of the first approximation, we present the set of equations (\ref{Eq07a}) and (\ref{Eq08}) in the second approximation on the parameter $\varepsilon$ as:
\begin{widetext}%
	\begin{eqnarray} %
	\frac{d\gamma }{dT} &=& \mp\frac{320 \beta }{\Delta_{f}^4}e^{2\theta} \int\limits_{0}^{\infty } \frac{q^{K+1}} {\left(e^{2\theta } + q^{K} \right)^{2} } \frac{q^{\pm 1} - 2}{\left(q +1\right)^{4} } dq, \label{ErmStep_Eq_20a}\\%
	\frac{d\theta}{dT} &=& \Delta V\gamma + 4\beta \gamma ^{3} \mp \frac{10 \Delta V^{2}}{27 \beta \gamma K^{4}}
	\int\limits_{0}^{\infty}\frac{e^{2\theta}\left(1 + 2\theta - K\ln{q}\right) - q^{K}}{\left(e^{2\theta} + q^{K}\right)^{2}}\frac{q^{\pm 1} - 2}{\left(q + 1\right)^{4}}\,q^{K+1}\,dq, \label{ErmStep_Eq_22}%
	\end{eqnarray}
\end{widetext}
where the upper and lower signs correspond to the upper and lower signs in the forcing function (\ref{Eq16}).

\begin{figure}[h!]
	\centerline{%
		\includegraphics*[width=1.7in]{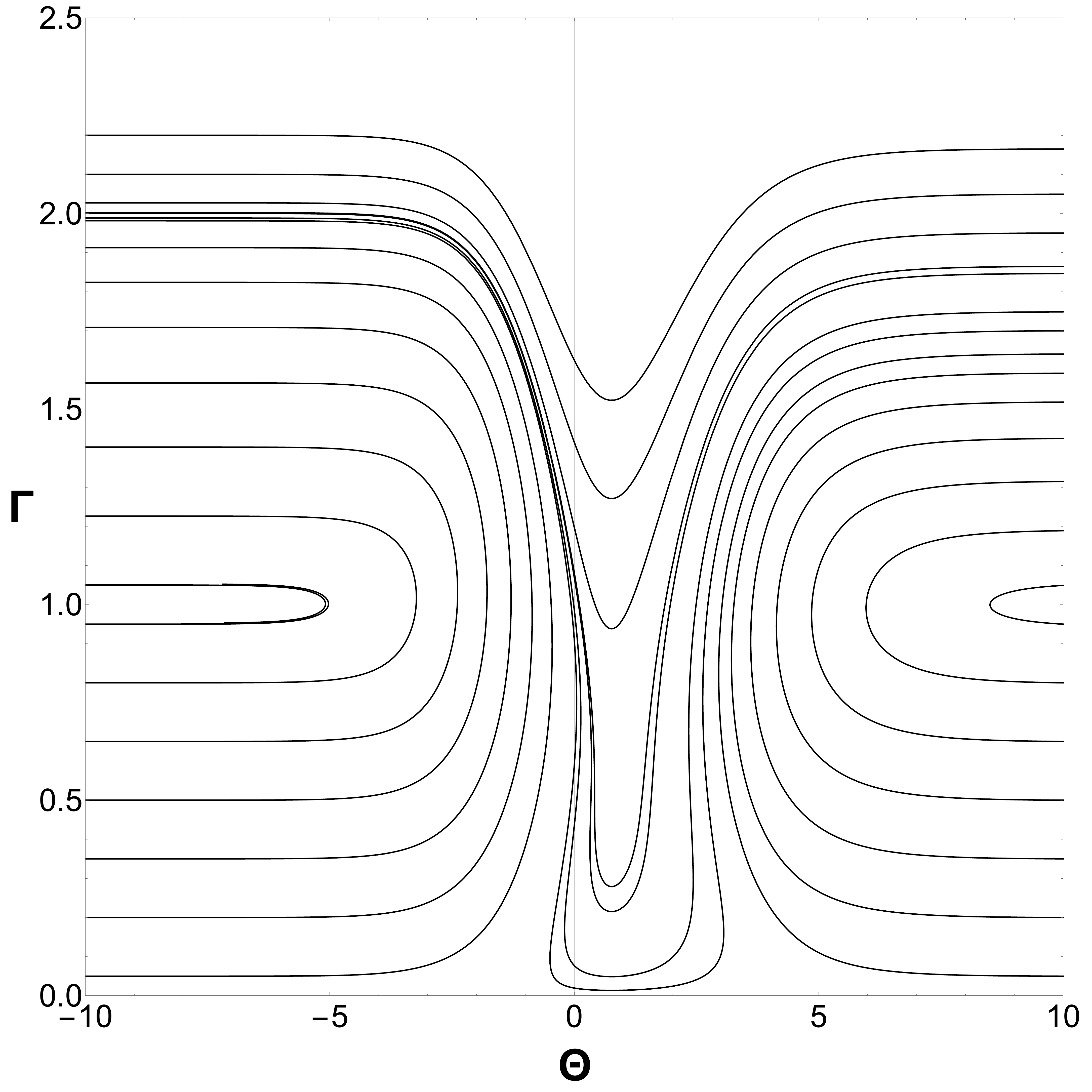}%
		\includegraphics*[width=1.7in]{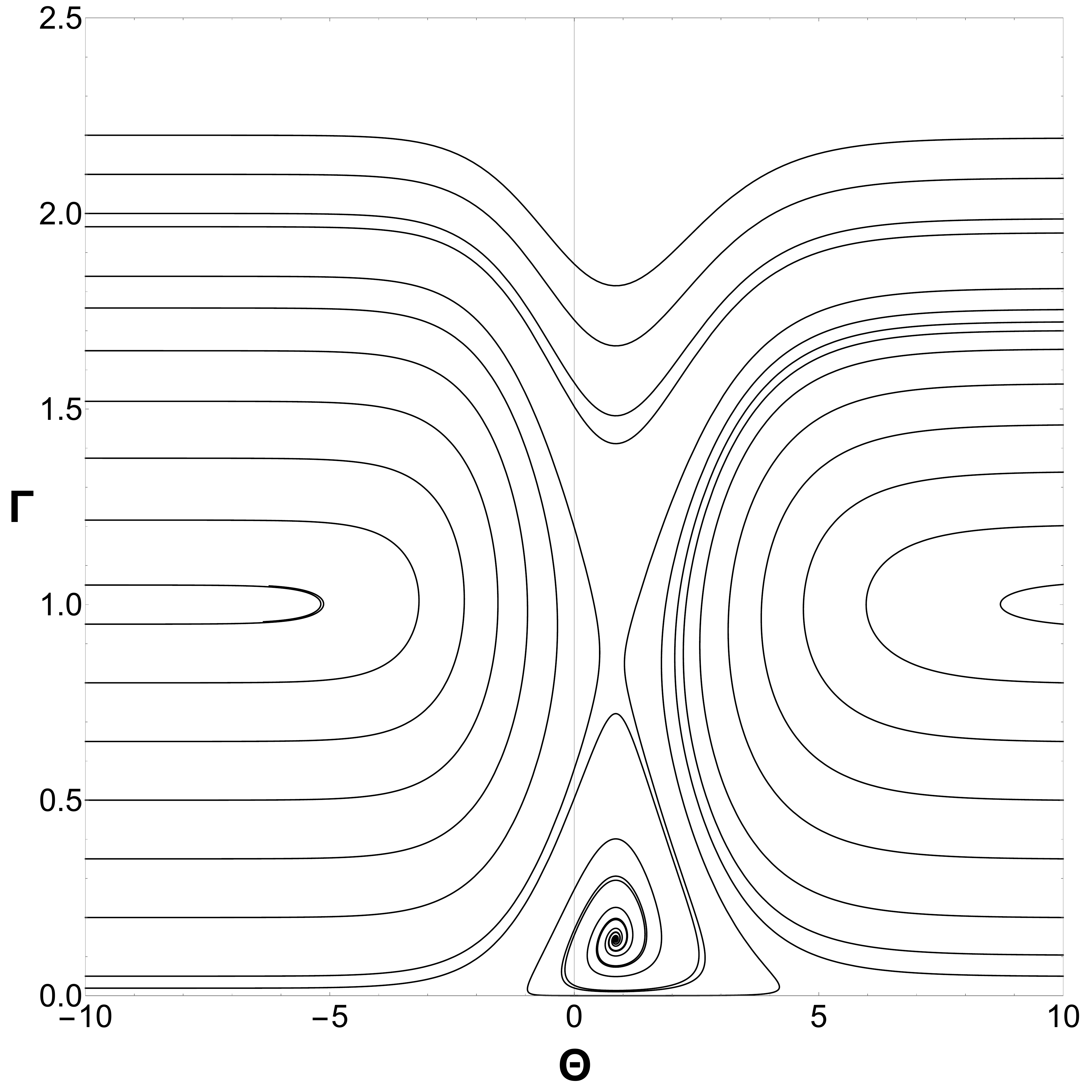}} %
	{\hspace{0.5cm} a) \hspace{3.8cm} b)}
	\caption{The phase portraits of the dynamical system (\ref{ErmStep_Eq_20a}) and (\ref{ErmStep_Eq_22}) with $K = 2$ (frame a) and $K = 3.5$ (frame b).}
	\label{ErmStep_Fig_6B}
\end{figure}

The set of equations (\ref{ErmStep_Eq_20a}) and (\ref{ErmStep_Eq_22}) does not have equilibrium states for relatively small width of the forcing $K \le 3$ as shown in Fig. \ref{ErmStep_Fig_6B}a). In the phase plane there are either transient trajectories or bouncing trajectories in this case. If the forcing width increases and becomes greater than $K > 3$, then the equilibrium state of a stable focus appears, which corresponds to the trapped KdV soliton of a small amplitude within the potential well as shown in Fig. \ref{ErmStep_Fig_6B}b). But when the forcing width further increases and becomes greater than $K > 5$, then the equilibrium state disappears again, and the phase portrait of the system (\ref{ErmStep_Eq_20a}) and (\ref{ErmStep_Eq_22}) becomes qualitatively similar to that shown in Fig. \ref{ErmStep_Fig_2}b).

\section{The Gardner-type forcing}
\label{Case3}

Consider now the forcing function in the form:
\begin{equation} %
\label{ErmStep_Eq_24A}%
f(x) = \frac{1}{\left[1 + B\cosh{\left(x/\Delta_f\right)}\right]^3},
\end{equation}
where $B$, and $\Delta_f$ are constant parameters. Its derivative is:
\begin{equation} %
\label{ErmStep_Eq_24B}%
f'_x(x) = -\frac{3 B (1+B)^3 \sinh{\left(x/\Delta_f\right)}}{ \Delta_f \left[1 + B\cosh{\left(x/\Delta_f\right)}\right]^4},
\end{equation}

For any parameters $\varepsilon$ and $\Delta_f$ this forcing provides the exact solution to the fKdV Eq. (\ref{Eq02}) in the form of Gardner soliton (see, e.g., Ref. \cite{OPSS2015}):
\begin{equation} %
\label{ErmStep_Eq_24}%
u\left(x\right) = \frac{A_f}{1 + B\cosh{\left(x/\Delta_f\right)}},
\end{equation}
where $A_f = 6\beta/\alpha\Delta_f^{2}$, $V = c + \beta/\Delta_f^{2}$. The parameters $B$ and $\Delta_f$ determine the amplitude of external force $\varepsilon$ by means of the formula:
\begin{equation} %
\label{ErmStep_Eq_29A}
\varepsilon = -\frac{12 \beta^2 (B-1)}{\alpha (B+1)^2 {\Delta_f}^4}.
\end{equation}
\begin{figure}[b]
	\centerline{\includegraphics[width=3.8in]{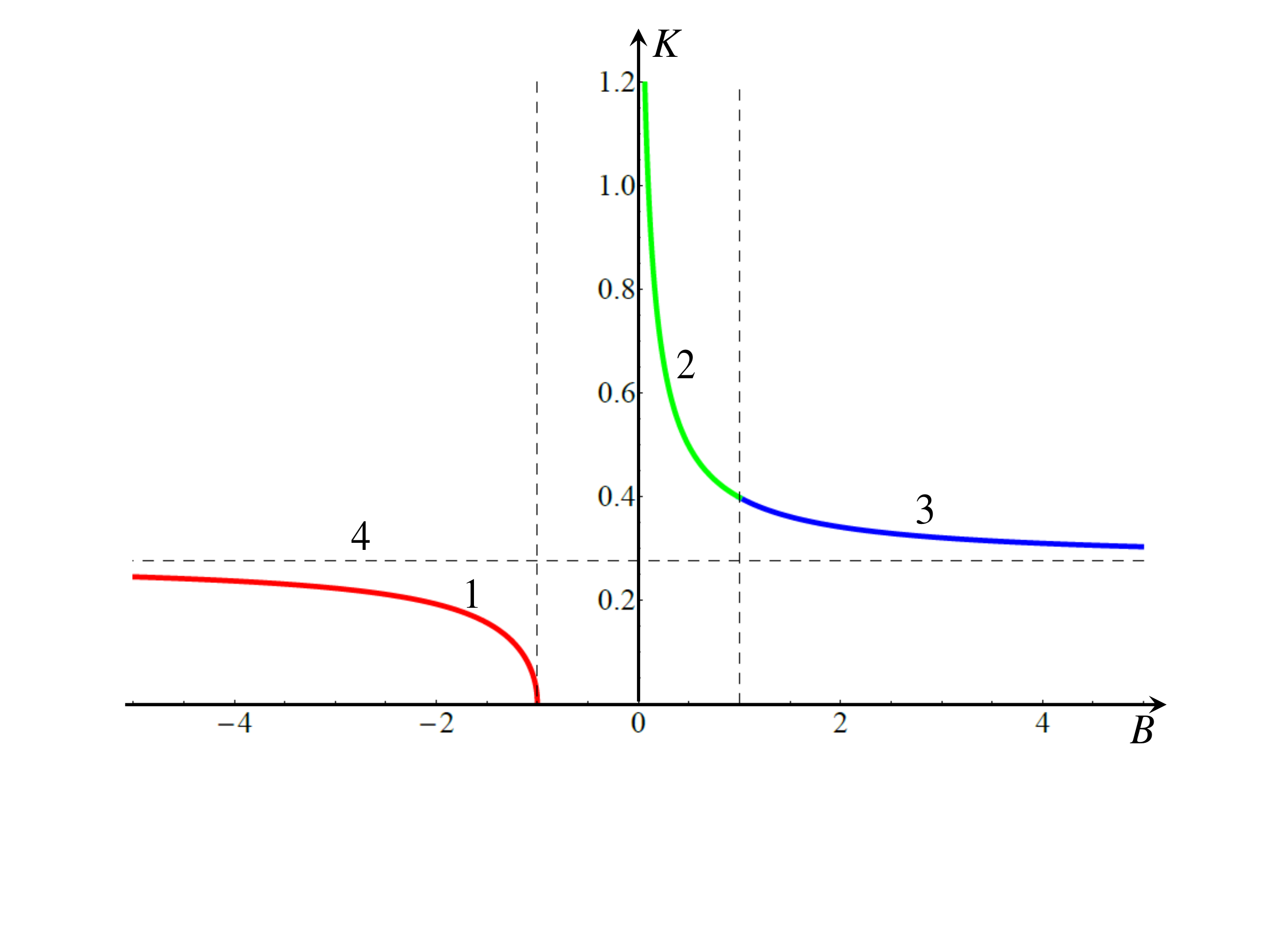}}
	\vspace*{-1.5cm}%
	\caption{The dependence of parameter $K$ characterising the relative width of forcing (\ref{ErmStep_Eq_24A}) as function of parameter $B$. Horizontal line 4 shows the asymptotic value of $K = \ln{3}/4 \approx 0.275$ when $B \to \pm \infty$.} %
	\label{ErmStep_Fig_7}
\end{figure}

Real nonsingular soliton solutions exist only for $B > 0$ and $B < -1$. When $B$ ranges from 0 to 1, we have a family of solitons varying from a KdV soliton, when $B \to 1_-$, to a table-top soliton, when $B \to 0_+$. When $1 \le B < \infty$, we obtain a family of bell-shaped solitons of positive polarity, and when $-\infty < B < -1$, -- a fa\-mily of bell-shaped solitons of negative polarity (see, e.g., Ref. \cite{OPSS2015}). The half-width of forcing (\ref{ErmStep_Eq_24A}), i.e. half-distance between the extrema of function $f'_x$, is determined by the parameter $B$:
\begin{equation} %
\label{ErmStep_Eq_29}
D_f = \Delta_f\ln{\left[\frac{1 \pm R \pm \sqrt{2\left(R + R^2 - 42B^2\right)\vphantom{2^2}}}{6B} \right]},
\end{equation}
where $R = \sqrt{1 + 48B^2}$, upper signs pertain to $B > 0$, and lower signs -- to $B < 0$. Figure \ref{ErmStep_Fig_7} shows the parameter $K = \gamma_0 D_f$ as the function of $B$.

\begin{figure}[t!]
\centerline{\includegraphics*[width=1.7in]{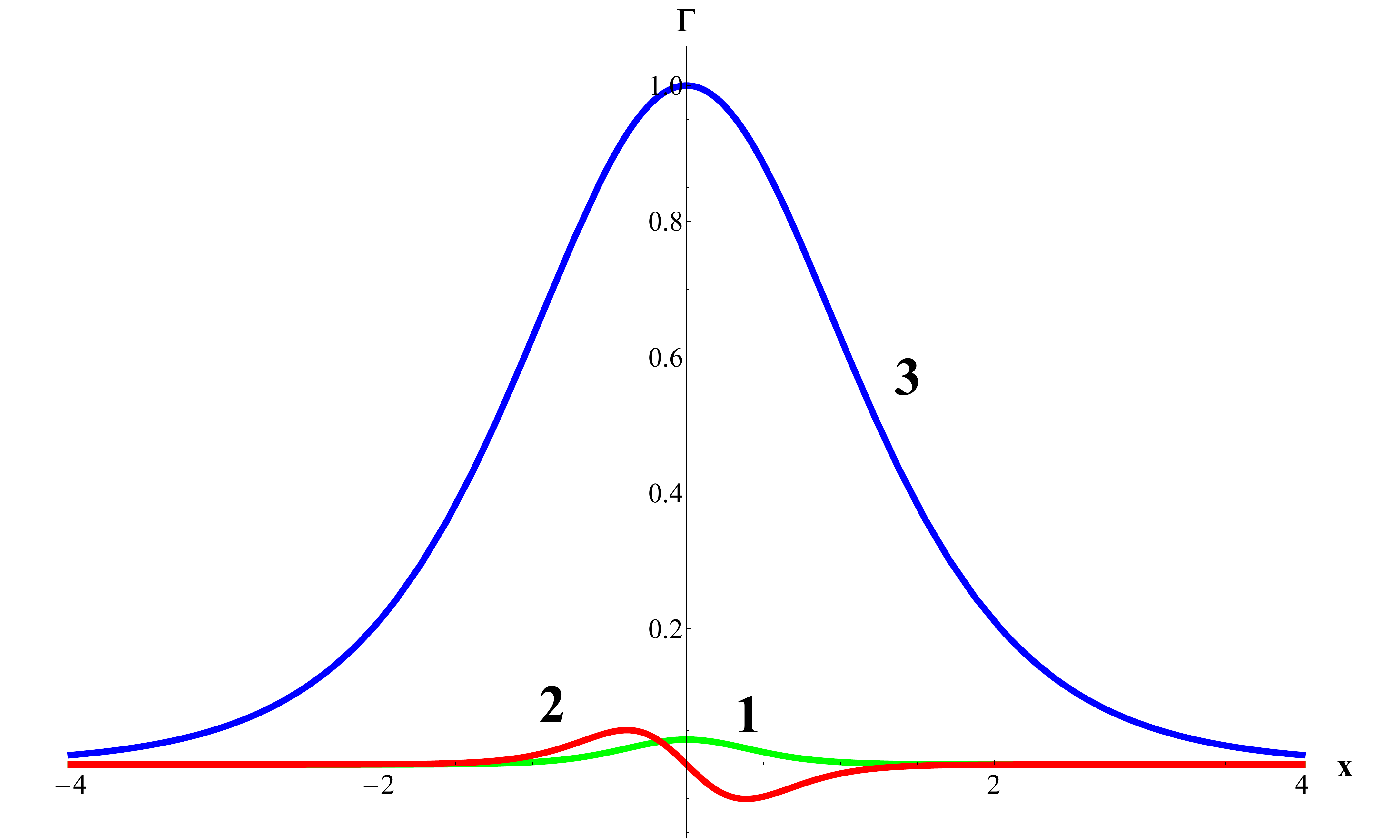} %
\includegraphics*[width=1.7in]{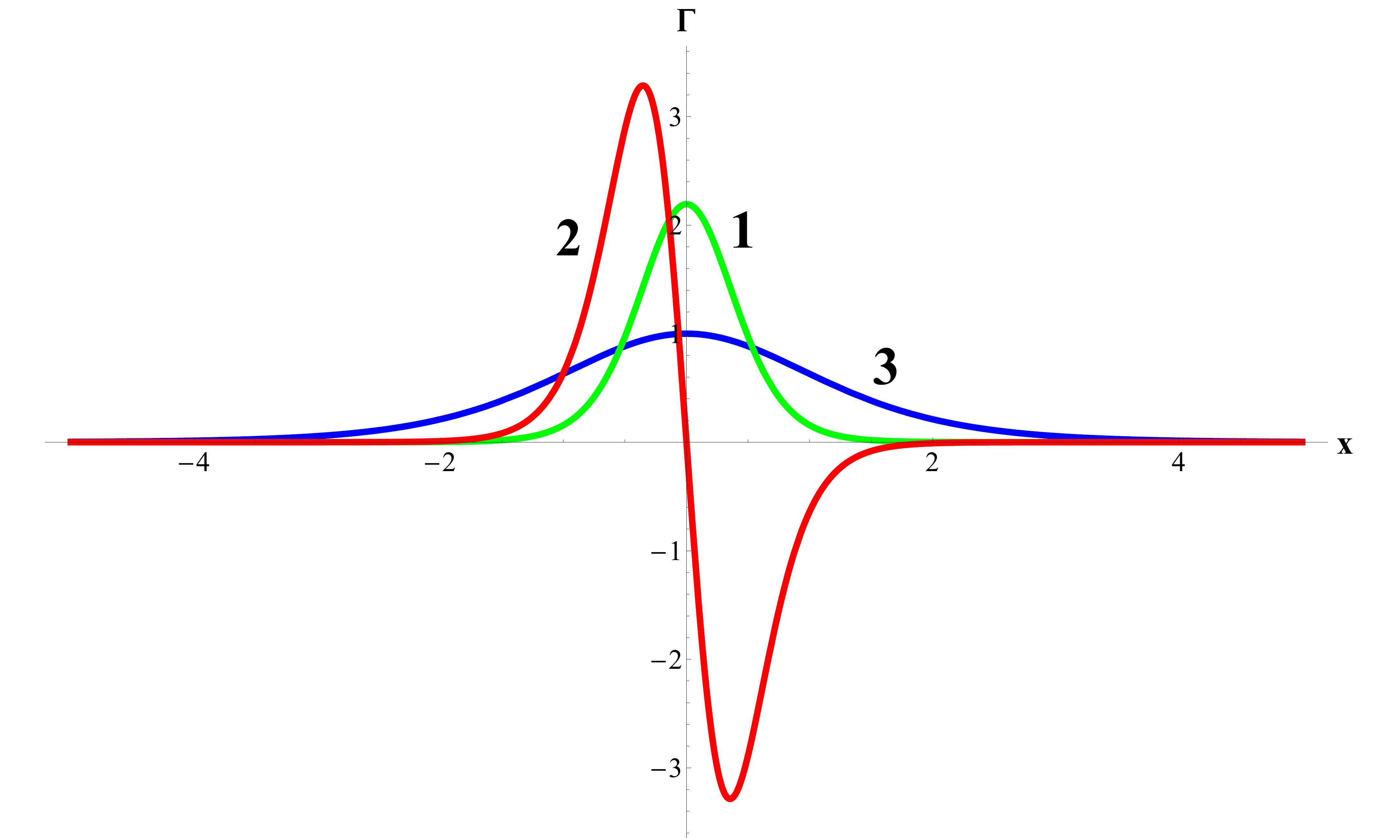}} %
\vspace{0.3cm} %
{\large a) \hspace{2.5cm} \phantom{aaaaa} b)}%
\caption{Green lines 1 represent the forcing function $f(x)$ as per Eq. (\ref{ErmStep_Eq_24A}), red lines 2 represent its derivatives $f'(x)$, and blue lines 3 show the initial KdV soliton (\ref{Eq03}). In frame a) $K = 0.274$, $B = -221.23$; in frame b) $K = 0.25$, $B = -6.08$.} %
\label{ErmStep_Fig_8}%
\vspace{1cm} %
\centerline{\includegraphics*[width=1.7in]{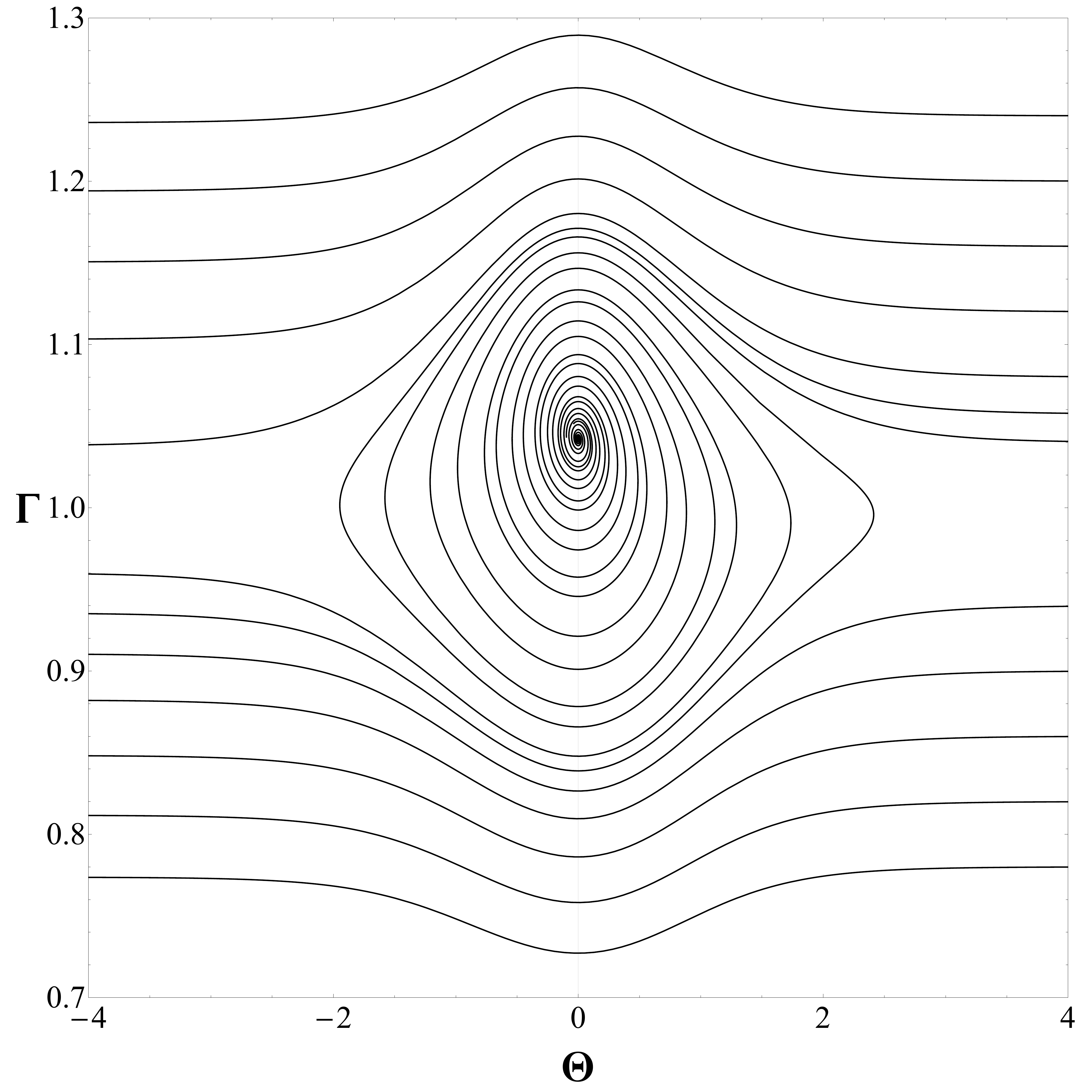}
\includegraphics*[width=1.7in]{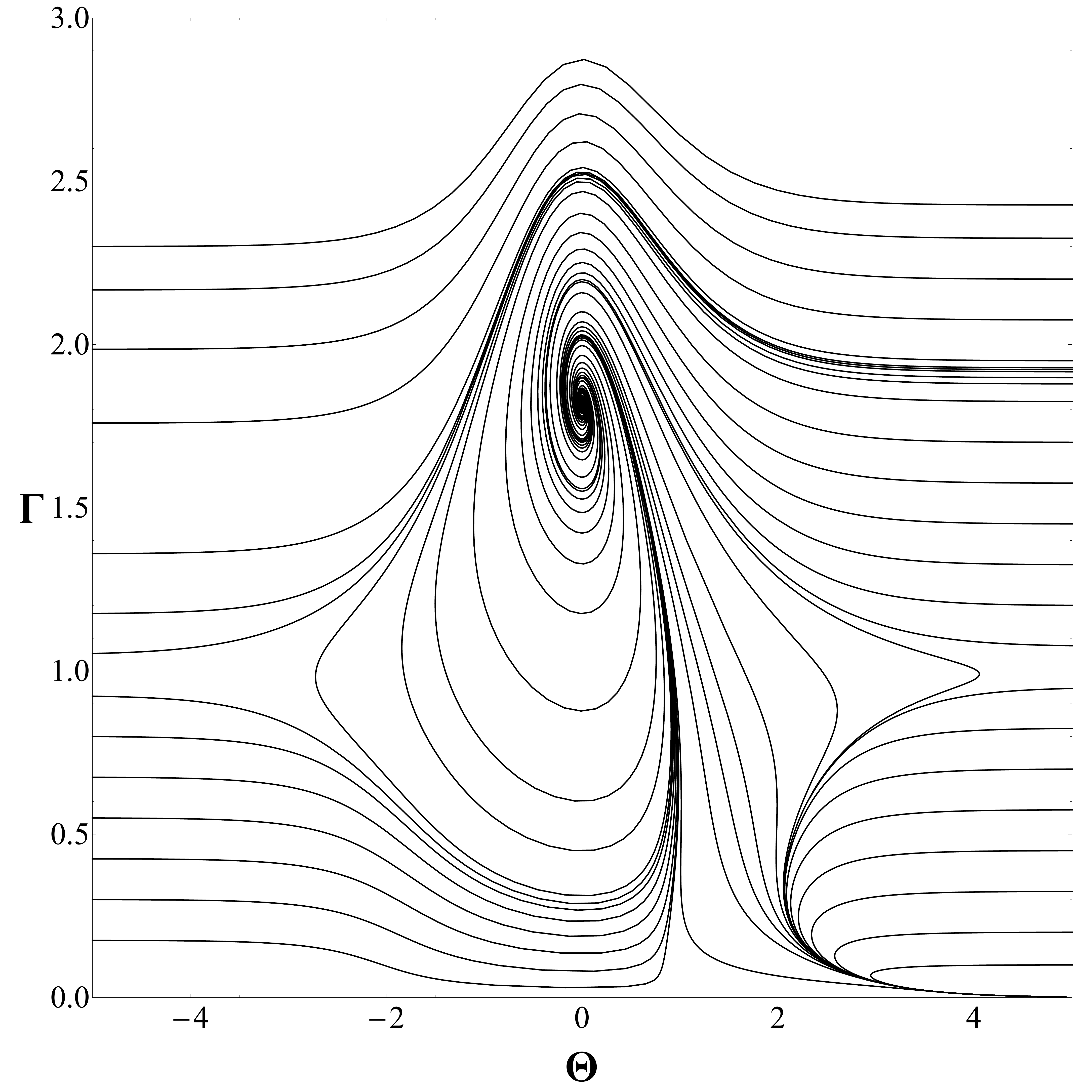}} %
\vspace{0.3cm} %
{\large \hspace*{0.1cm} a) \hspace{3.8cm} b)}%
\caption{Phase portraits of the dynamical system (\ref{ErmStep_Eq_26a}) and (\ref{ErmStep_Eq_28}) for $K = 0.274$ (frame a); and $K = 0.25$ (frame b).}
\label{ErmStep_Fig_10}
\end{figure}

In the interval $-\infty < B < -1$ the forcing is narrow, $K < 1$ (see line 1 in Fig. \ref{ErmStep_Fig_7}), and function $f(x)$ is positive (see green line 1 in Fig. \ref{ErmStep_Fig_8}a). In the interval $1 < B < \infty $ the forcing is narrow too (see line 3 in Fig. \ref{ErmStep_Fig_7}), but function $f(x)$ is negative (see green line 1 in Fig. \ref{ErmStep_Fig_11}a). In the interval $0 < B < 1$ (see line 2 in Fig. \ref{ErmStep_Fig_7}) the forcing can be both wide, $K > 1$, when $B$ is very close to zero, and narrow, $K < 1$, in the rest of this interval; the forcing function is positive within the entire interval $0 <B \le 1$ (see green lines 1 in Fig. \ref{ErmStep_Fig_14}. Note that as follows from Eq. (\ref{ErmStep_Eq_29A}), the amplitude of forcing vanishes when $B \rightarrow \pm \infty$, and we have a KdV soliton of arbitrary amplitude freely moving without external action. When $B \rightarrow -1_-$, the forcing width becomes zero, but its amplitude goes to infinity; the forcing looks like the Dirac $\delta$-function.

When $B \rightarrow 0_+$, the forcing becomes infinitely wide. These two limiting cases have been studied in Ref. \cite{GPT1994}, and our purpose here is to study the situations when $K$ is of the order of unity.

Assume again that the amplitude of external force is small $\varepsilon \ll 1$ and the initial perturbation has the form of KdV soliton (\ref{Eq03}) moving with the initial velocity $\upsilon_0 = V$. This gives $\gamma_0\Delta_f = 1/2$. After substitution of function $f'_x(x)$ from Eq.~(\ref{ErmStep_Eq_24B}) and soliton solution (\ref{Eq03}) into Eqs. (\ref{Eq07a}) and (\ref{Eq08}) and denoting $p = e^{\Phi/\Delta_f}$, we obtain in the second approximation the following set of equations:
\begin{figure}[h!]
\vspace*{0.5cm} %
\centerline{\includegraphics*[width=1.7in]{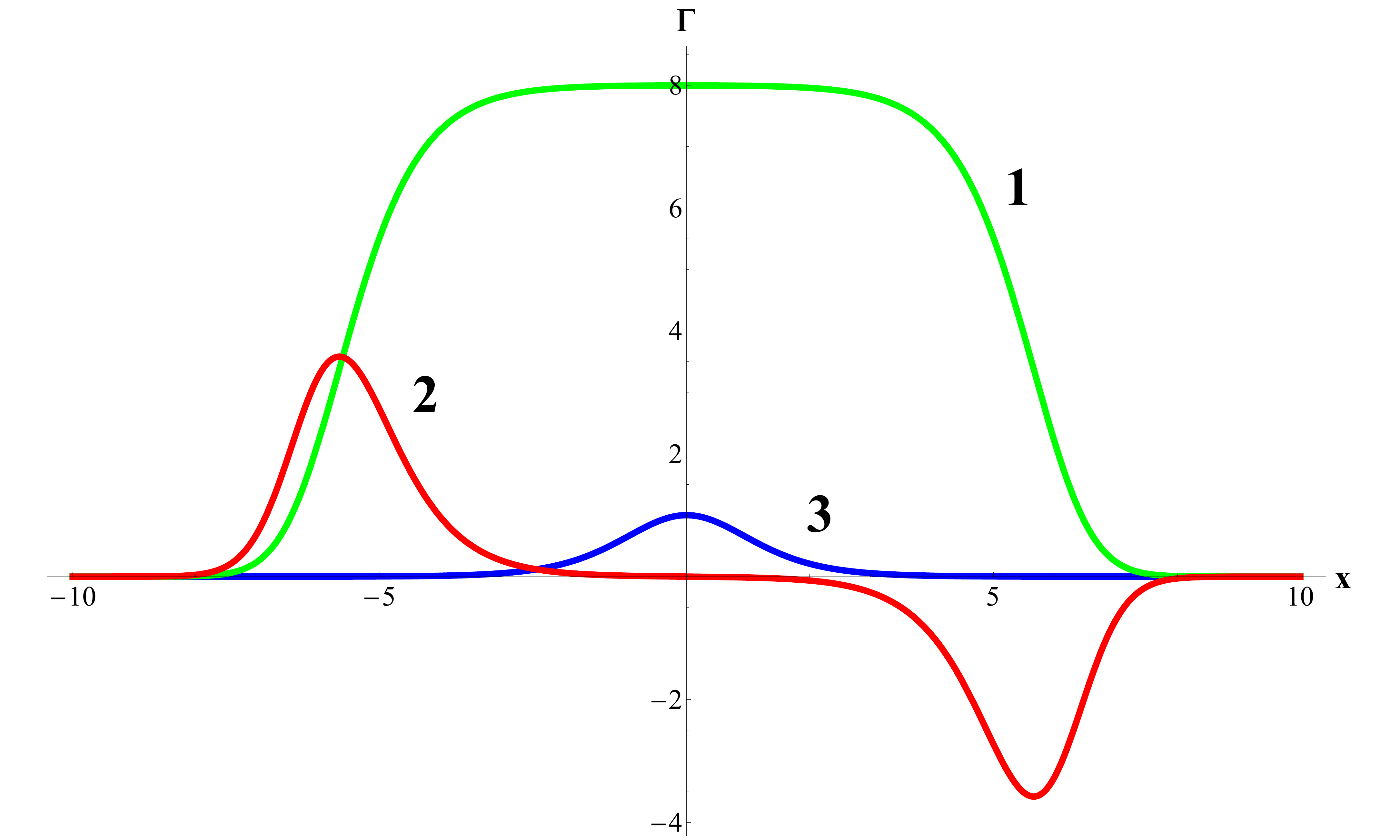}%
\includegraphics*[width=1.7in]{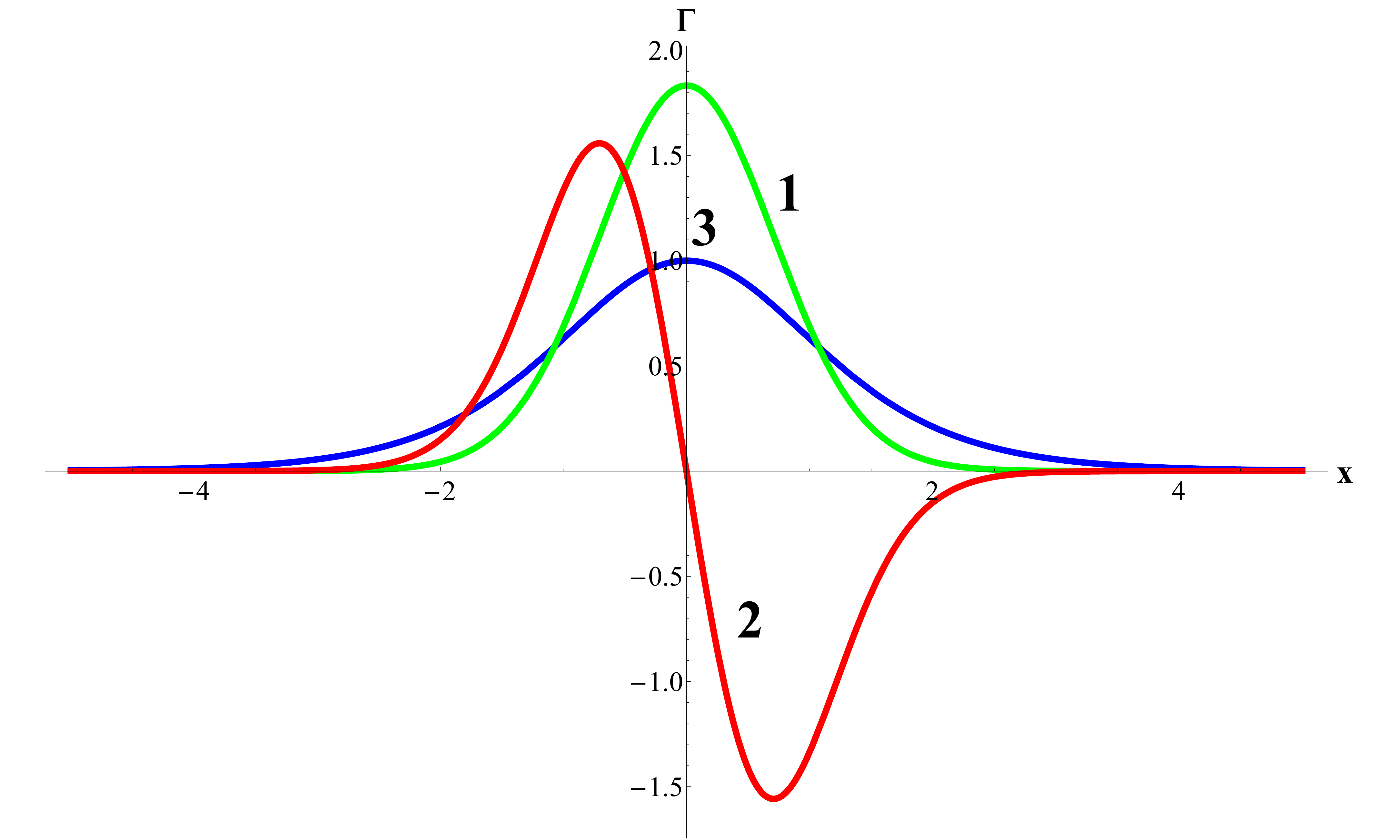}} %
	\vspace{0.3cm} %
	{\large a) \hspace{2.5cm} \phantom{aaaaa} b)}%
	\caption{Green lines 1 represent the forcing function $f(x)$ as per Eq. (\ref{ErmStep_Eq_24A}), red lines 2 are its derivatives $f'(x)$, and blue lines 3 are the initial KdV soliton (\ref{Eq03}). In frame a) $K = 2$, $B = 0.012$; in frame b) $K = 0.5$, $B = 0.49$.} %
	\label{ErmStep_Fig_14}%
	\vspace{1cm} %
	\centerline{\includegraphics*[width=1.7in]{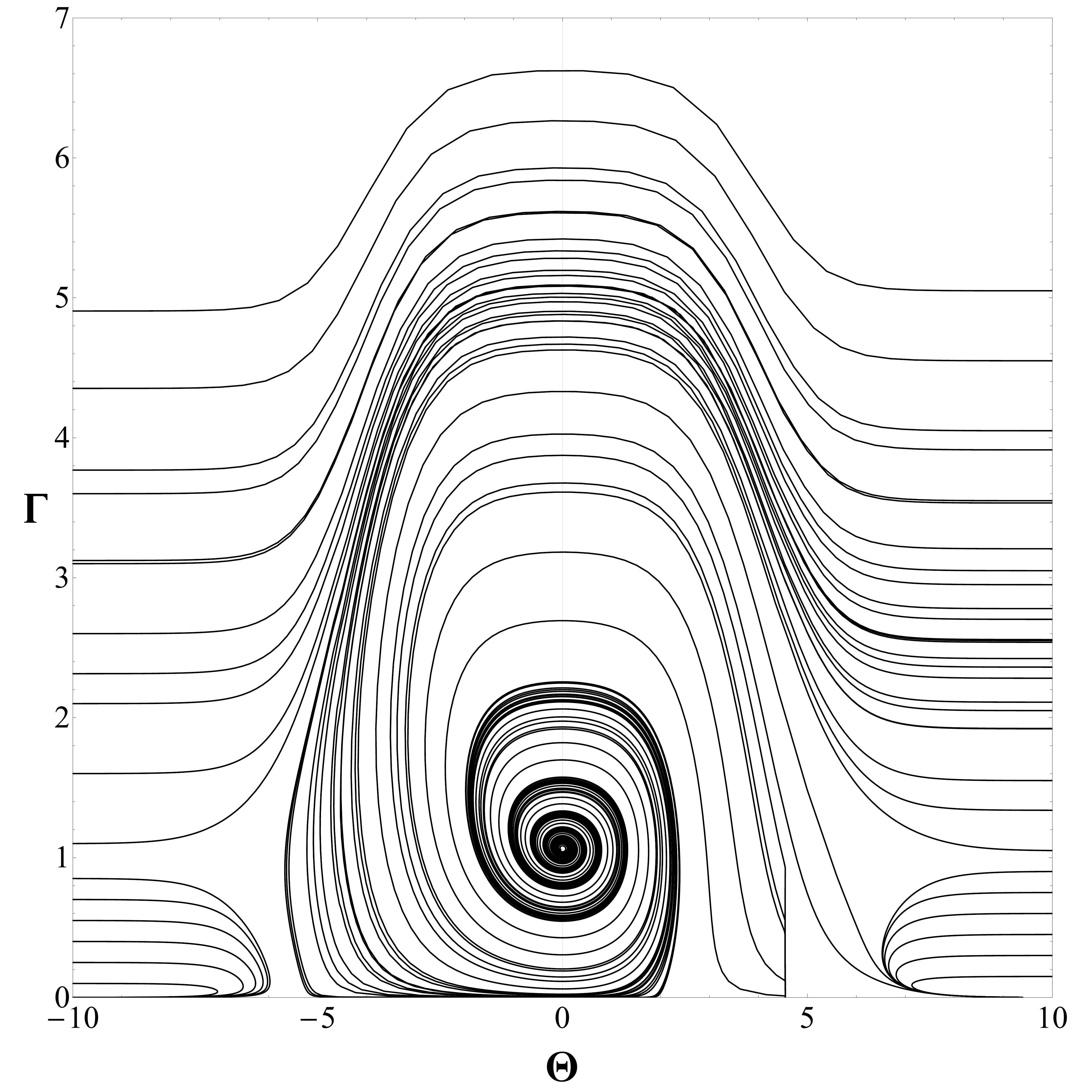}%
		\includegraphics*[width=1.7in]{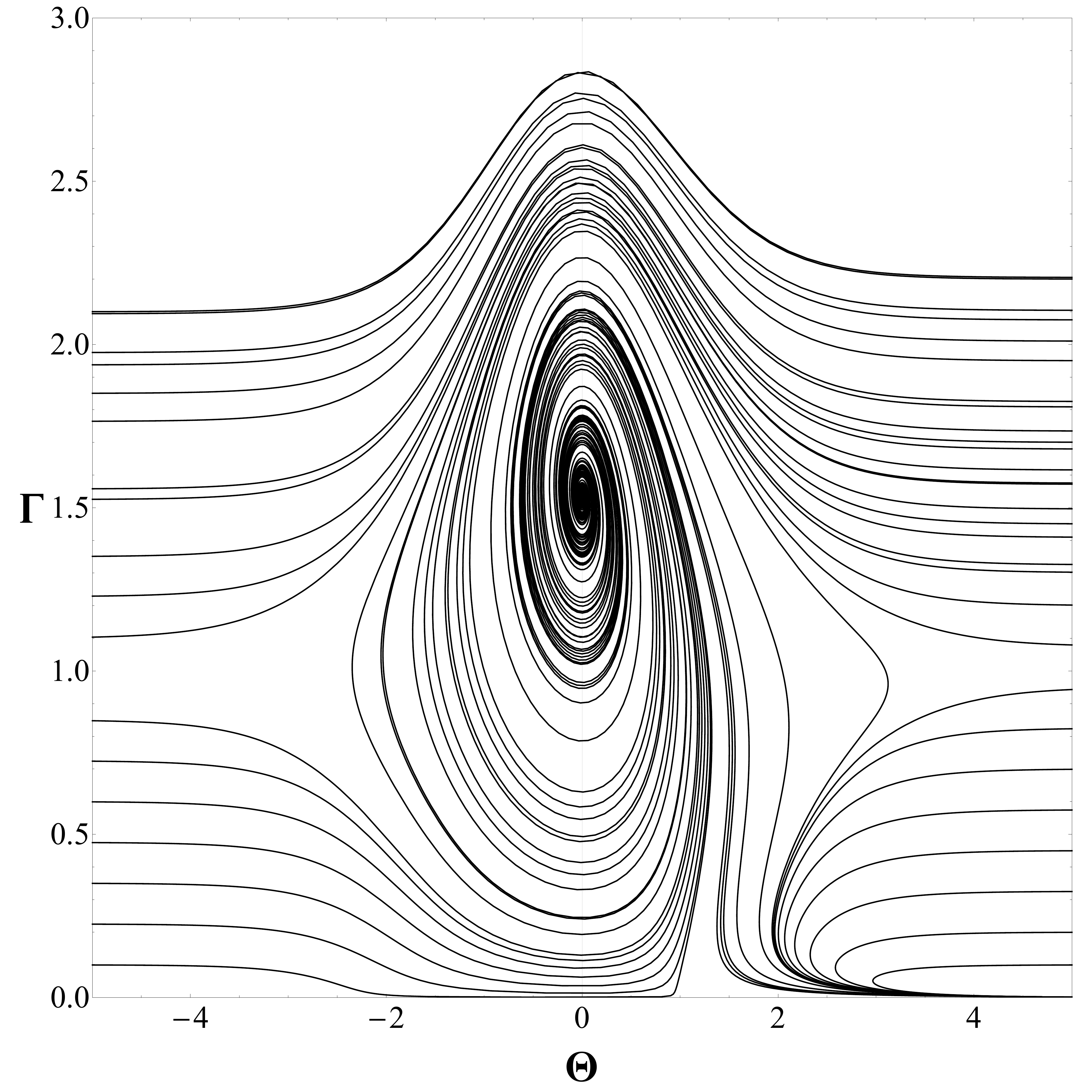}}%
\vspace{0.2cm} %
	\hspace*{0.2cm} {\large a) \hspace{2.0cm} \phantom{aaaaaaa} b)}%
	\caption{Phase portraits of the dynamical system (\ref{ErmStep_Eq_26a}) and (\ref{ErmStep_Eq_28}) for $K = 2$ (frame a); and $K = 0.5$ (frame b).}
	\label{ErmStep_Fig_16}
\end{figure}
\begin{figure}[h!]
	\centerline{\includegraphics*[width=1.7in]{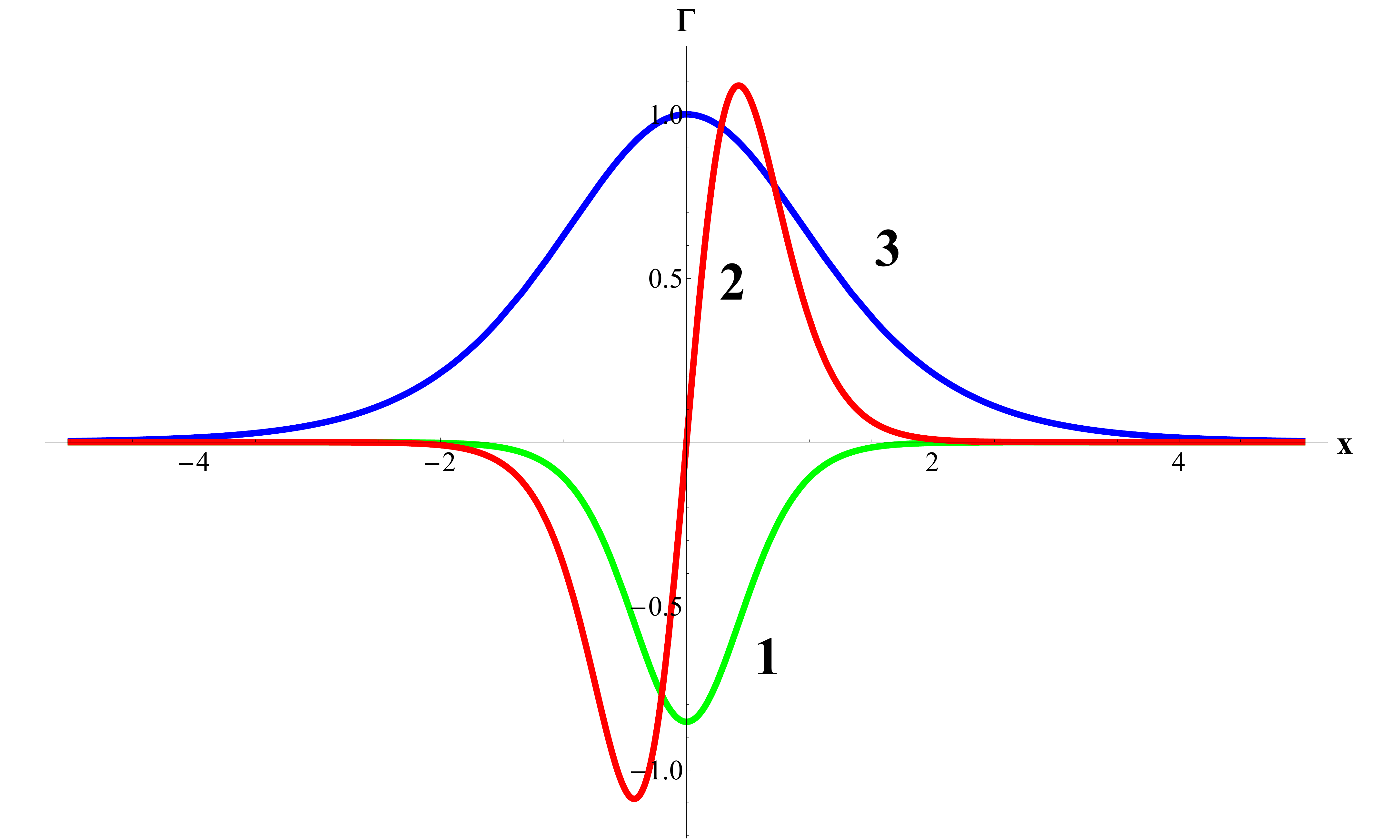}%
		\includegraphics*[width=1.7in]{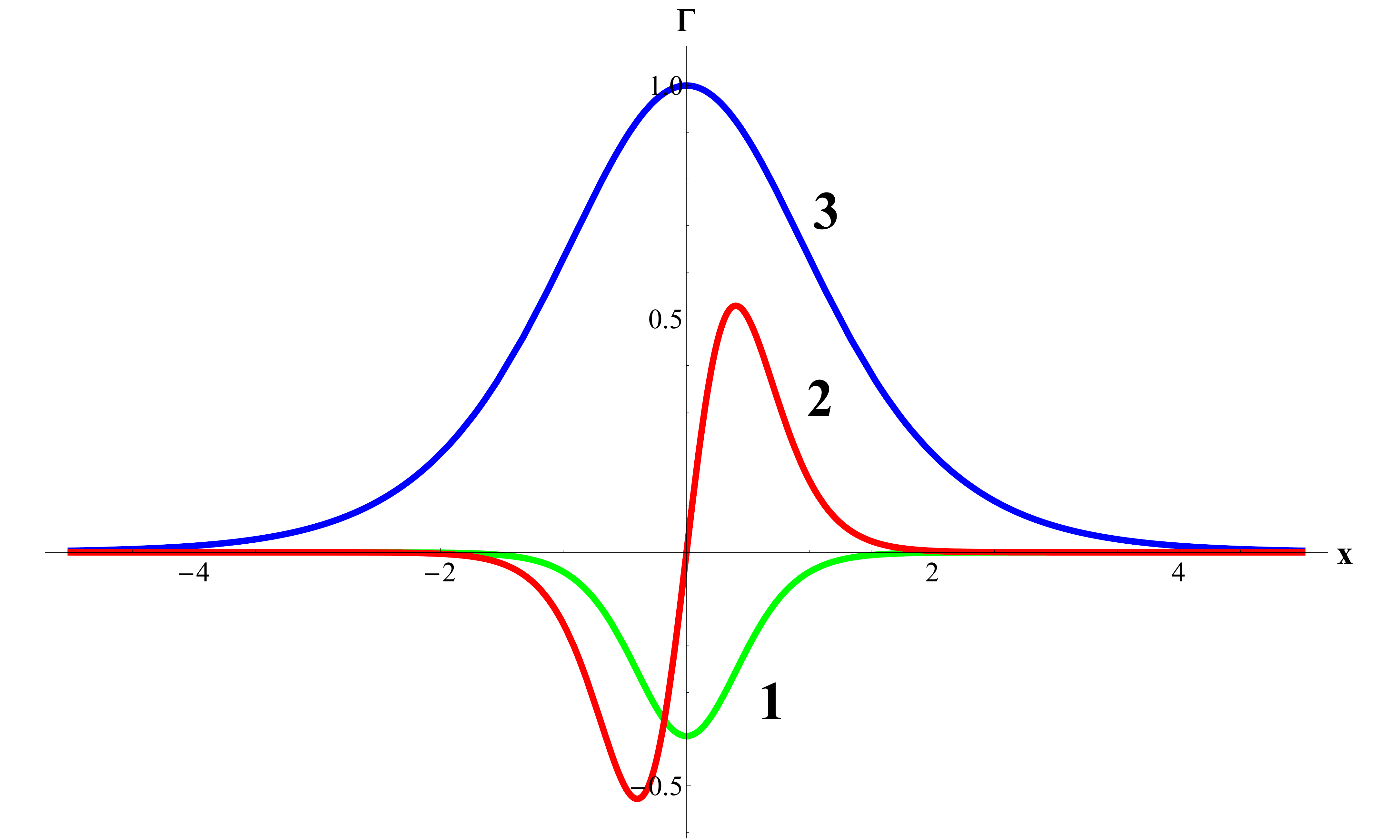}} %
	\vspace{0.2cm} %
	{\large a) \hspace{2.5cm} \phantom{aaaaa} b)}%
	\caption{Green lines 1 are the forcing functions $f(x)$ as per Eq. (\ref{ErmStep_Eq_24A}), red lines 2 are their derivatives $f'(x)$, and blue lines 3 are the initial KdV solitons (\ref{Eq03}). In frame a) $K = 0.3$, $B = 5.5$; in frame b) $K = 0.283$, $B = 17$.} %
	\label{ErmStep_Fig_11}%
\end{figure}

\begin{widetext}%
	\begin{eqnarray} %
		\frac{d\gamma }{dT} &=& \frac{48 (B^2 - 1) \Delta V^2}{\beta B^3}e^{2\theta}\int\limits_{0}^{\infty } \frac{p^{3}\left(p^{2} - 1\right) dp} {\left(e^{2\theta} + p\right)^{2}\left(p^2 + 2p/B + 1\right)^{4}}, \label{ErmStep_Eq_26a}\\%
		&& \nonumber \\
		\frac{d\theta }{dT} &=& \Delta V\gamma + 4\beta \gamma ^{3} - \frac{24\Delta V^{2}}{\beta\gamma} \frac{B^{2} - 1}{B^{3}}
		\int\limits_{0}^{\infty} \frac{e^{2\theta}\left(1 + 2\theta - \ln{p}\right) - p}{\left(e^{2\theta} + p\right)^{2}} \frac{p^3\left(p^{2} - 1\right)dp}{\left(p^2 + 2p/B + 1\right)^{4}}. \label{ErmStep_Eq_28}
	\end{eqnarray}
\end{widetext}%

Below we describe the changes in the phase portraits of the dynamical system (\ref{ErmStep_Eq_26a}) and (\ref{ErmStep_Eq_28}) when the parameter $B$ varies from minus to plus infinity. When this para\-meter is negative, $-\infty < B < -1$, the forcing is narrow $K < 1$ and positive (see lines 1 in Fig. \ref{ErmStep_Fig_8}). Such forcing with a hump cannot trap a soliton, therefore there is only one equilibrium state, the unstable focus (alias the unstable spiral), which implies that a soliton placed at this state escapes it under the action of infinitely small perturbations (see Fig. \ref{ErmStep_Fig_10}). The only difference between the portraits shown in Figs. \ref{ErmStep_Fig_10}a) and \ref{ErmStep_Fig_10}b) is that there are no transient trajectories in the latter figure below the equilibrium point, but instead the bouncing trajectories appear in the right lower corner.

When the parameter $B$ varies in the range $0 < B \le 1$, the forcing can be both wide, $K > 1$, and narrow, $K < 1$, but in both cases the potential function is positive (see lines 1 in Fig. \ref{ErmStep_Fig_14}). Again, due to the positive hump-type forcing incapable to trap a soliton, the only one equilibrium state on the phase plane is possible, the unstable focus. The typical phase portraits in this case are qualitatively similar both for the wide and narrow forcing (cf. Figs. \ref{ErmStep_Fig_16}a) and \ref{ErmStep_Fig_16}b) for $K = 2$, $B = 0.012$ and $K = 0.5$, $B = 0.49$ respectively).

\begin{figure*}%
	\centerline{\includegraphics*[width=7in]{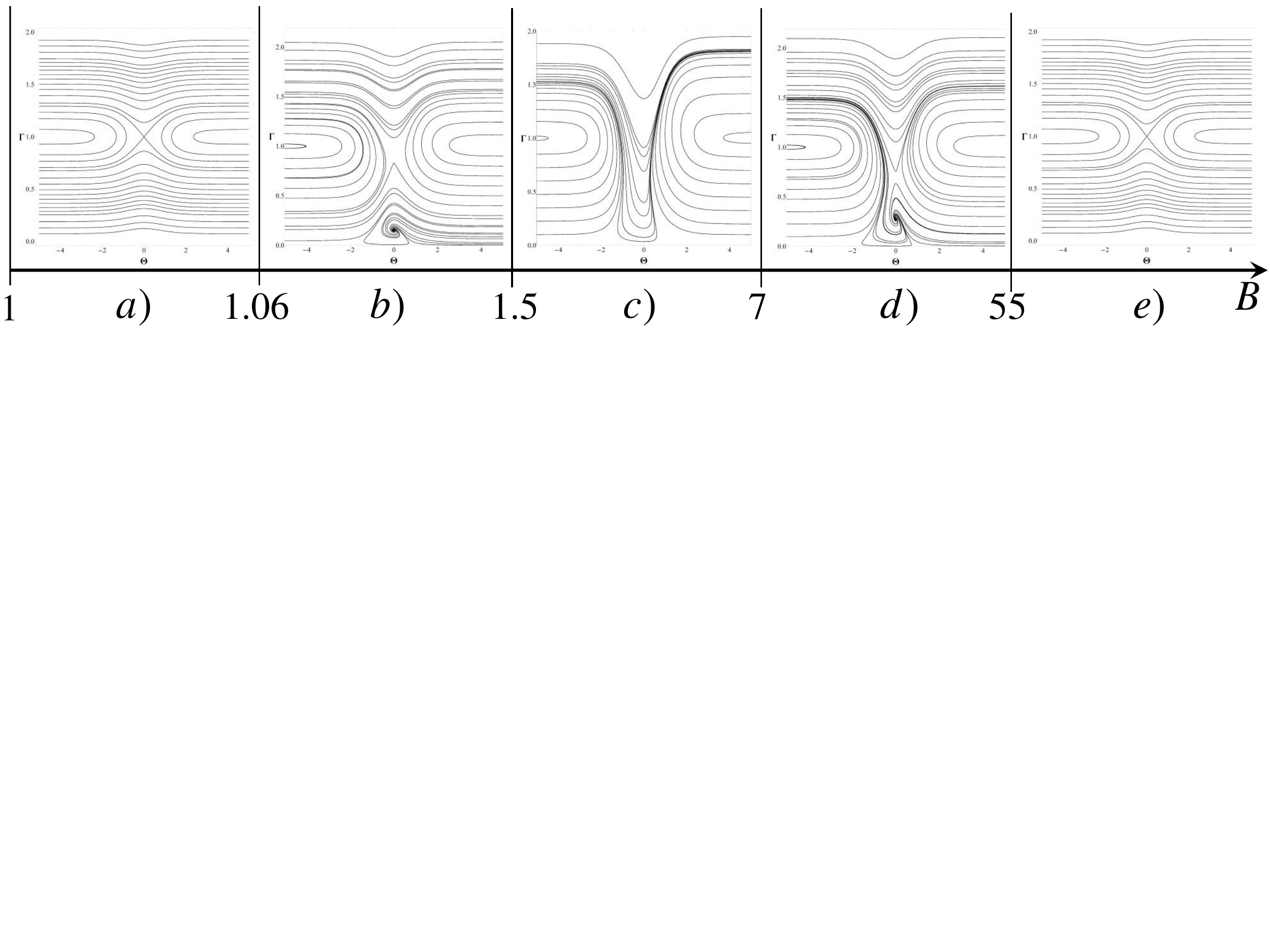}} %
	\vspace*{-8.5cm}
	\caption{
		Phase portraits of the system (\ref{ErmStep_Eq_26a}) and (\ref{ErmStep_Eq_28}) in the different interval of parameter $B \ge 1$ corresponding to the narrow forcing ($0.275 < K < 0.397$), of a negative polarity. The portraits were generated for the following parameters: frame a): $B = 1.05$, $K = 0.392$; frame b): $B = 1.2$, $K = 0.379$; frame c): $B = 3$, $K = 0.32$; frame d): $B = 10$, $K = 0.288$; frame e): $B = 60$, $K = 0.277$.
	}
	\label{ErmStep_Fig_13}
\end{figure*}

When $B > 1$, the forcing is narrow $0.275 < K < 1$, but now negative (see lines 1 in Fig. \ref{ErmStep_Fig_11}). Such forcing with a well can trap a soliton of a very small amplitude in the certain intervals of parameter $B$. In the interval $1 \le B < B_1 (\approx 1.06)$, there is only one unstable equilibrium state of a saddle type; the typical phase portrait is shown in Fig. \ref{ErmStep_Fig_13}a). Then, in the interval $B_1 < B < B_2 (\approx 1.5)$ there is an equilibrium state of the stable focus type; the corresponding phase portrait is shown in frame b). In the next interval $B_2 \le B < B_3 (\approx 7)$ the equilibrium state disappears, and the typical phase portrait is shown in frame c). In the interval $B_3 < B < B_4 (\approx 55)$ an equilibrium state of the stable focus type appears again; the corresponding phase portrait is shown in frame d). And at last, in the interval $B > B_4$ the unstable equilibrium state of a saddle type like in the frame a) arises again (see frame e). In the latter case forcing amplitude becomes very small (it asymptotically vanishes when $B \to \infty$), therefore it is incapable to retain a soliton.

\section{A periodic forcing}
\label{PeriodForce}

Consider now soliton dynamics in the nonstationary external field periodically varying in time and space. A similar problem has been studied in Refs. \cite{Grimshaw1994,GPS1996,Cabral2004,Cox2005}. As has been shown in those papers, a periodic forcing can lead to both dynamic and chaotic regimes of wave motion. Here we consider a model of forcing which generalises the model studied in Ref. \cite{GPS1996} and admits exact solutions. In contrast with the Ref. \cite{GPS1996}, we do not use here again the approximation of either soliton or external forcing by the Dirac delta-function and study only the dynamic behaviour of a soliton in a periodically varying forcing.

Let us assume that in Eq. (\ref{Eq01}) the forcing function has the form:
\begin{equation} %
\label{Eq17a}%
f(x, t) = \sigma F(t)\sech^2{\left[\frac{x - \int S(t)dt}{\Delta_f}\right]},%
\end{equation} %
where $F(t)$ and $S(t)$ are arbitrary functions of their argument, and $\sigma$ is a real parameter.

As has been shown in Ref. \cite{LZM2001}, the fKdV equation (\ref{Eq01}) with such forcing function has the exact solution for any parameters $\varepsilon$ and $\Delta_f$ in the form of a soliton moving with the variable velocity $S(t)$:
\begin{equation} %
\label{Eq18}%
u(x, t) = \frac{12\beta}{\alpha\Delta_f^2} \sech^2{{\left[\frac{x - \int S(t)dt}{\Delta_f}\right]}},
\end{equation} %
where $\sigma = 12\beta/\left(\varepsilon\alpha\Delta_f^2\right)$, and $S(t) = c + 4\beta/\Delta_f^2 - F(t)$.

Let us choose, in particular,
\begin{equation} %
\label{Eq18F}
F(t) = \frac{\varepsilon\alpha}{12\beta}\Delta_f^2\left(1 + \tilde{V}\sin{\varepsilon\omega t}\right),
\end{equation} %
where $\tilde{V}$ and $\omega$ are arbitrary real parameters; then solution (\ref{Eq18}) represents a soliton moving with the mean velocity $V$ as per Eq. (\ref{Eq09}) and periodically varying component proportional to $\tilde{V}\cos{(\varepsilon\omega t)}$:
\begin{equation} %
\label{Eq091} %
V_{tot} = c + \frac{4\beta}{\Delta_f^2} - \frac{\varepsilon \alpha\Delta_f^2}{12\beta}\left(1 + \tilde{V}\sin{\varepsilon\omega t}\right).
\end{equation}

With the choice of $F(t)$ as above, the forcing function has the same shape as in Fig. \ref{ErmStep_Fig_1}, but now the amplitude of for\-cing function $f(x,t)$ periodically varies in time and the forcing moves with periodically varying speed. Note that in Ref. \cite{GPS1996} the authors considered variation of only forcing phase, whereas in our case both the forcing amplitude and phase vary in time.

If $\varepsilon \ll 1$ is a small parameter as above and the initial perturbation has the form of a KdV soliton (\ref{Eq03}), then from the slightly modified asymptotic theory described in Section \ref{Case1} we obtain very similar equations for the first- and second-order approximations. To show this, let us make the transformation of independent variables in Eq. (\ref{Eq01}) $\hat{x} = x - \int S(t)dt$, $\hat{t} = t$, then Eq.~(\ref{Eq01}) reduces to the form similar to Eq. (\ref{Eq02}) (the symbol $\hat{~}$ is further omitted):
\begin{equation} %
\label{Eq022}%
\frac{\partial u}{\partial t} +[c-S(t)]\frac{\partial u}{\partial x}
+\alpha _{} u\frac{\partial u}{\partial x} +\beta \frac{\partial
	^{3} u}{\partial x^{3} } = \varepsilon\frac{df(x)}{d x}.
\end{equation}

In the presence of small external perturbation solitary wave solution (\ref{Eq03}) gradually varies, and its amplitude, half-width
$\gamma^{-1}$, and velocity become slow functions of time $T = \varepsilon t$, so that the soliton phase can be determined as in Section \ref{Sect-2}: $\Phi = x - \Psi(T)$ (cf. Eq. (\ref{Eq04a})), but with the periodically varying speed:
\begin{equation}
\label{Eq041}%
\upsilon(T) = \frac{\alpha A(T)}{3} - \frac{4\beta}{\Delta_f^2} + \frac{\varepsilon\alpha}{12\beta}\Delta_f^2\left(1 + \tilde{V}\sin{\omega T}\right).
\end{equation}

The time dependence of soliton amplitude follows from the energy balance equation (\ref{Eq06}). Then carrying out the asymptotic analysis up to the second order on the para\-meter $\varepsilon$, we eventually obtain the set of equations similar to Eqs. (\ref{Eq11}) and (\ref{Eq19}) with the only modifications caused by the periodic factor in front of integrals:

\begin{figure*}[t!]
	\centerline{\includegraphics*[width=7in]{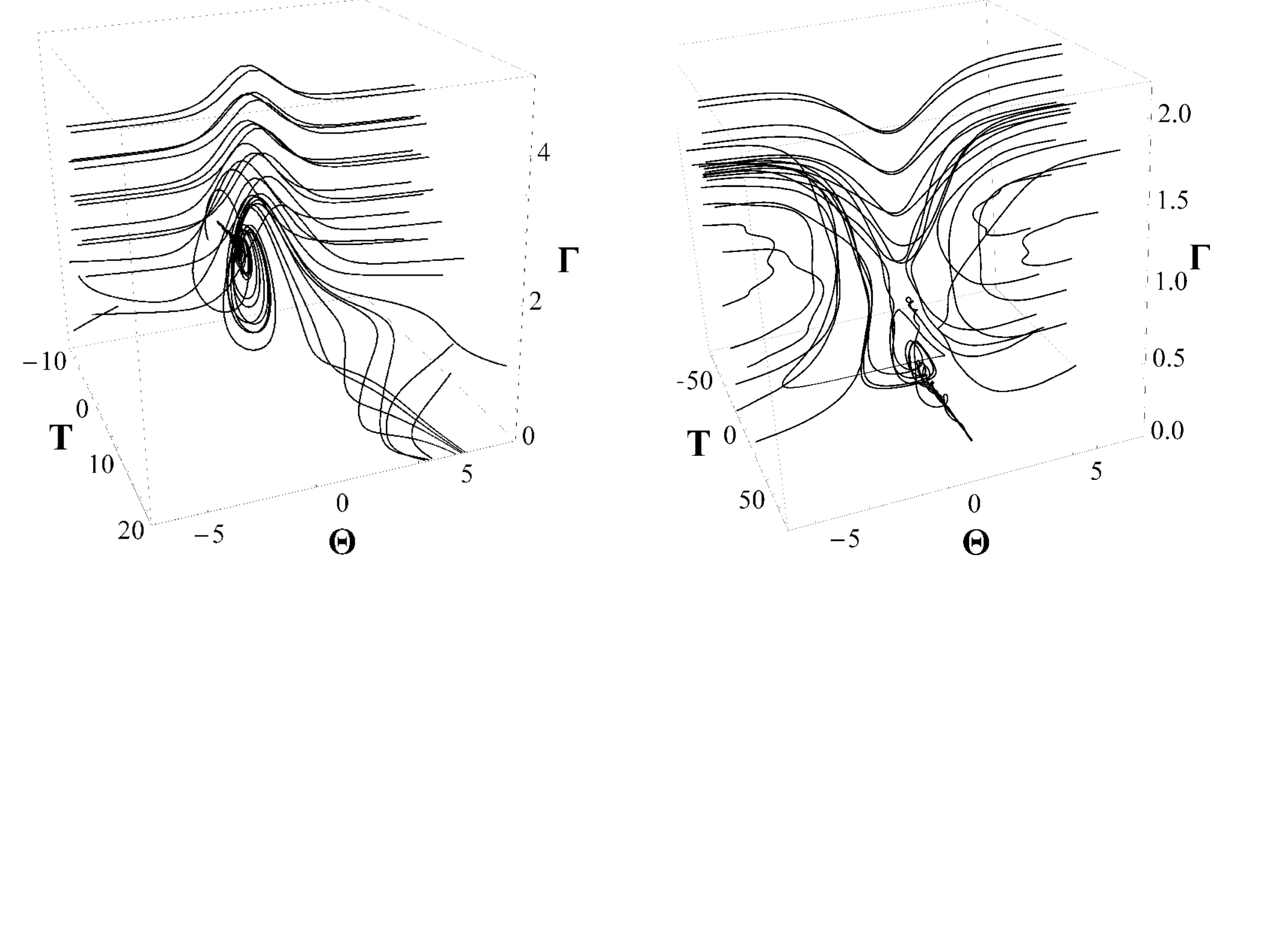}} %
	\vspace{-5cm}%
	{\large a) \hspace{2.5cm} \phantom{aaa} b)}%
	\caption{The phase space $\theta, \Gamma, T$ of the non-stationary dynamical system (\ref{NonStat1}), (\ref{NonStat2}).}
	\label{ErmStep_Fig_161}
\end{figure*}

\begin{widetext}
	\begin{eqnarray} %
	\frac{d \gamma }{d T} &=& \frac{2\varepsilon\alpha}{3 \beta}\left(1 + \tilde{V}\sin{\omega T}\right)e^{2\theta} \int \limits_0^{\infty } \frac{q^{K}}{\left(e^{2\theta} + q^{K}\right)^2}\frac{q - 1}{\left(q + 1\right)^3}dq, \label{NonStat1}\\
	\frac{d \theta}{dT}& = &\Delta V(T) \gamma +4\beta \gamma ^3 - \frac{\varepsilon\alpha }{3\beta\gamma } \left(1 + \tilde{V}\sin{\omega T}\right) \int \limits_0^{\infty} \frac{e^{2 \theta} \left(1 + 2\theta - K\ln{q}\right) -  q^{K}}{\left(e^{2\theta} + q^{K}\right)^2}\frac{q - 1}{\left(q + 1\right)^3} q^{K}dq, \label{NonStat2}
	\end{eqnarray} %
\end{widetext}
where now
$$
\Delta V(T) = c - V_{tot} = -\frac{4\beta}{\Delta_f^2} + \frac{\varepsilon\alpha}{12\beta}\Delta_f^2\left(1 + \tilde{V}\sin{\omega T}\right).
$$

There are no analytical solutions to this set of equations, but it can be solved numerically, and a qualitative character of solutions can be illustrated by means of three-dimensional phase space, where the third coordinate is the $T$-axis. Few typical phase trajectories are shown in Fig.~\ref{ErmStep_Fig_161} for the positive and negative forcing functions (cf. with the phase plane shown in Fig.~\ref{ErmStep_Fig_3}). Due to oscillations of forcing functions, the phase trajectories revolve around the unstable (frame a) or stable (frame b) focus-type equilibria and displace along the $T$-axis. Trajectories in frame (a) eventually become parallel to the $T$-axis; this corresponds to solitary waves escaping from the forcing and uniformly moving with the constant amplitudes and speeds. Trajectories in frame (b), in contrast, eventually converge to the equilibrium point corresponding to the solitary wave trapped by the negative forcing. Such solitary wave ultimately moves synchronously with the forcing having periodically varying amplitude and speed.

\section{Results of numerical study}
\label{NumRes}

To validate the theoretical results obtained on the basis of asymptotic theory, we undertook direct numerical calculations within the framework of original forced KdV equation (\ref{Eq01}) with the different shapes of forcing $f(x, t)$. Below we present the most typical examples for the Gardner-type forcing considered in Section \ref{Case3}. In other cases the results obtained were qualitatively similar to presented here. Numerical solutions were obtained by means of the finite-difference code described in \cite{Obregon2012} and realised in Fortran.

First of all, it was confirmed that in all cases when the forcing is of positive polarity, there is no trapped soliton moving synchronously with the forcing. Even when a KdV soliton was placed initially at the centre of the hump-type forcing, it eventually escaped from the forcing and moved independently. A hump-type narrow forcing was capable to retain a KdV soliton only for while in agreement with the analytical prediction -- see the phase planes shown in Figs. \ref{ErmStep_Fig_3}a), \ref{ErmStep_Fig_10}, and \ref{ErmStep_Fig_16}b). In the case of a wide forcing the situation becomes more complicated and leads to the permanent generation of solitary waves at the rear slope of the forcing. Below we describe in detail soliton interaction with the wide and narrow forcing using as an example the Gardner-type forcing.

\begin{figure}[t!]
	\centerline{
		\includegraphics[width=11cm]{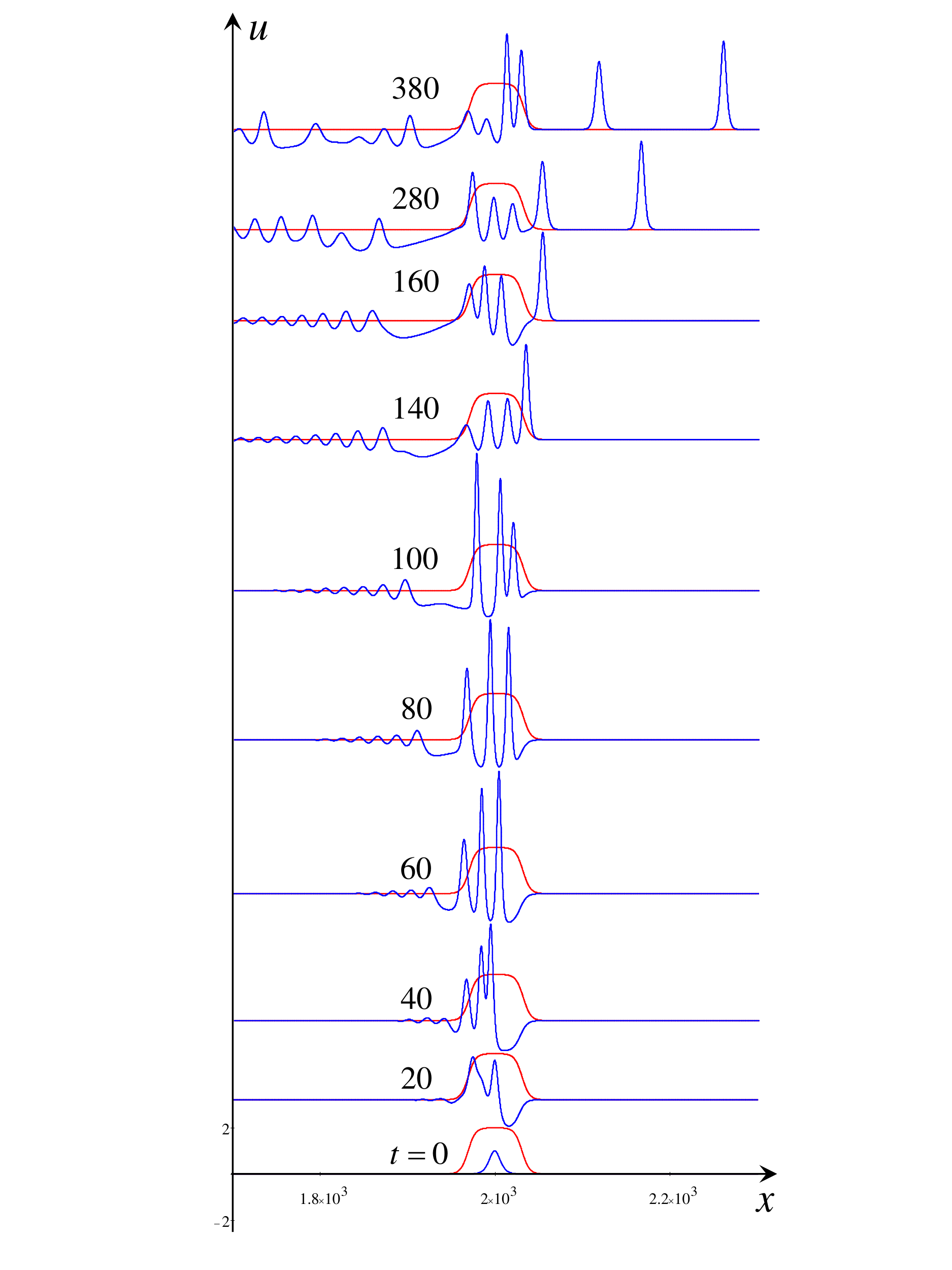}} %
	\vspace*{-0.5cm}
	\caption{Generation of solitons (blue lines) by wide forcing (red lines) in the case of Gardner-type forcing (only a fragment of the spatial domain of total length $4000$ is shown). The numerical solution of Eq. \ref{Eq01} was obtained with the following parameters $\alpha = 1$, $\alpha_1 = -0.5$, $\beta = 6$, $B = 0.012$.} %
	\label{Fig14}
\end{figure}

We considered solutions for $B > 0$ starting from small $B = 0.012$ when the forcing represents a $\Pi$-shaped pulse as shown in Fig. \ref{ErmStep_Fig_14}a). According to the asymptotic theo\-ry, such forcing leads to the unstable node/spiral on the phase plane (see Fig. \ref{ErmStep_Fig_16}a), which corresponds to the ge\-neration of solitons escaping from the forcing zone and mo\-ving to the right. However, when a soliton emerging from a small perturbation escapes from the forcing zone, ano\-ther soliton is created, and the process is repeated many times. Moreover, because the forcing is wide for such parameter $B$, several solitons can coexist within the forcing zone; then some of them leave this zone while new solitons are generated on the left slope of the forcing function. This was indeed observed, and results obtained are shown in Fig. \ref{Fig14}.

As one can see from this figure, the initial small-amplitude bell-shaped soliton at $t = 0$ starts to grow, but at the same time another perturbation generates on the left slope of the forcing function. Very quickly the number of solitons within the forcing reaches three, then one of them leaves the forcing zone at $t = 140$ and simultaneously one more small soliton is generated at the left slope of forcing. Then the second soliton leaves the forcing zone at $t = 280$ and the process repeats. Thus, the forcing acts as a generator of infinite series of random-amplitude solitons. The details of this and all subsequent processes can been seen in the videos available at the website \cite{Ermakov2018}.

\begin{figure}[h!]
	\centerline{\includegraphics[width=8cm]{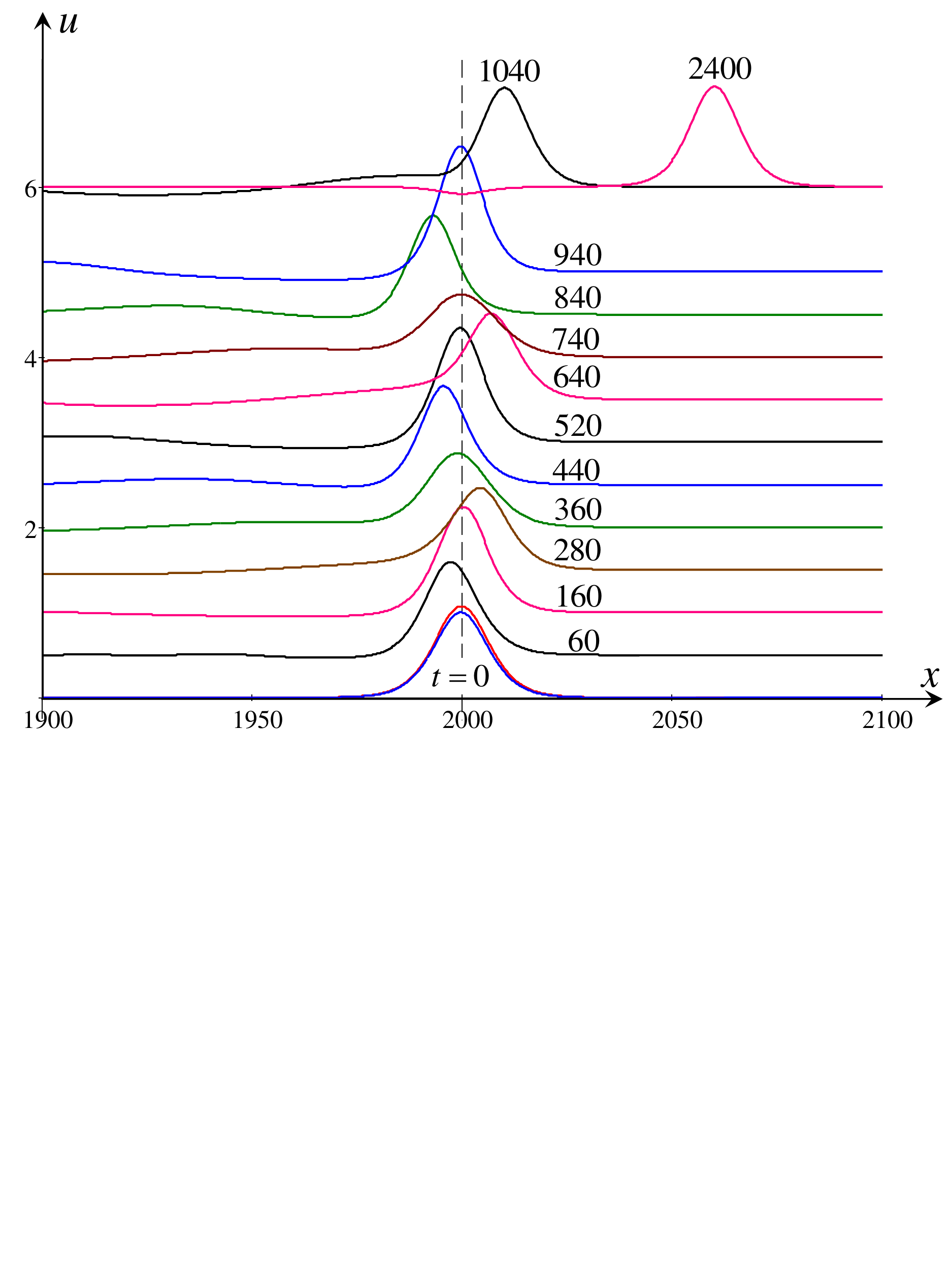}} %
	\vspace*{-4.5cm}
	\caption{Three oscillations of initial KdV soliton (blue line at $t = 0$) and its subsequent separation from the forcing zone at $t > 1040$. The Gardner-type forcing is shown by red line at $t = 0$; dashed vertical line shows the position of forcing maximum. The numerical solution was obtained with the following parameters of Eq. \ref{Eq01} $\alpha = 1$, $\alpha_1 = -0.125$, $\beta = 6$, $B = 0.85$, and $L = 4000$.} %
	\label{Fig15}
\end{figure}

When the forcing is relatively narrow, it can retain for a while only one soliton, which after a few oscillations within the forcing zone, eventually escapes and freely moves ahead. This is illustrated by Fig. \ref{Fig15}. In this figure one can see at $t = 0$ the KdV soliton (blue line) and the forcing (slightly taller pulse shown by red line). In the coordinate frame where the forcing is in the rest, the KdV soliton moves to the left first attaining the maximal deviation from the centre at $t \approx 60$; then it moves to the right attaining the maximal deviation from the centre at $t \approx 280$; then it moves again to the left, and so on. However, after three oscillations back and force, it leaves the forcing zone after $t = 1040$ and freely moves further as shown in the figure at $t = 2400$.
\begin{figure}[t!]
	\centerline{\hspace*{0.6cm}
		\includegraphics[width=10.6cm]{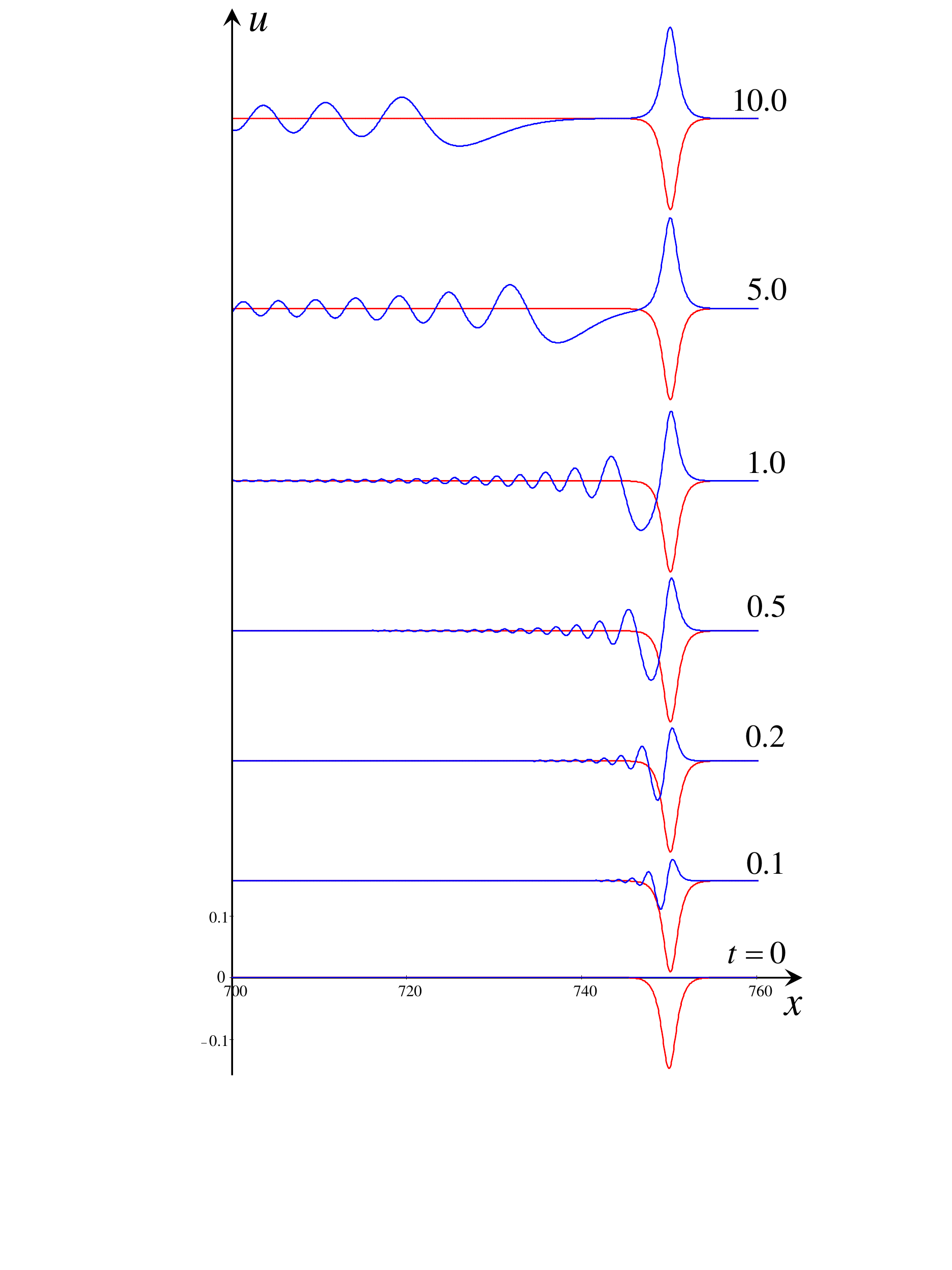}} %
	\vspace*{-2.2cm}
	\caption{Generation of a soliton (blue lines) by the negative forcing (red lines) from a random numerical noise.} %
	\label{Fig16}
\end{figure}

The situation is different when $B > 1$ and the forcing is negative (see Fig. \ref{ErmStep_Fig_13}). Among numerous situations arising in this case we shall describe here the most typical scenarios occurring at $B = 12.5$ and correspon\-ding to the phase plane shown in Fig. \ref{ErmStep_Fig_13}d). In this case there is a stable equilibrium state of the node type, which means that a soliton can emerge from small perturbations under the influence of a forcing. This was observed in numerical study with the zero initial condition as shown in Fig. \ref{Fig16} at $t = 0$ (all subsequent numerical calculations were obtained with the following parameters in Eq. (\ref{Eq01}): $\alpha = 6$, $\alpha_1 = 465.75$, $\beta = 1$, $B = 12.5$, and the total length of computational domain $L = 1500$). From a random numerical noise a perturbation grows within the forcing zone and becomes well visible at $t = 0.1$. Then it continues growing and developing into a soliton; this process is accompanied by emission of a quasi-linear wave train. Ultimately the wave train disappears, moving to the left and dispersing, whereas a soliton remains stable after being captured at the centre of the forcing in accordance with the theoretical prediction. 

\begin{figure}[h!]
	\centerline{\includegraphics[width=8cm]{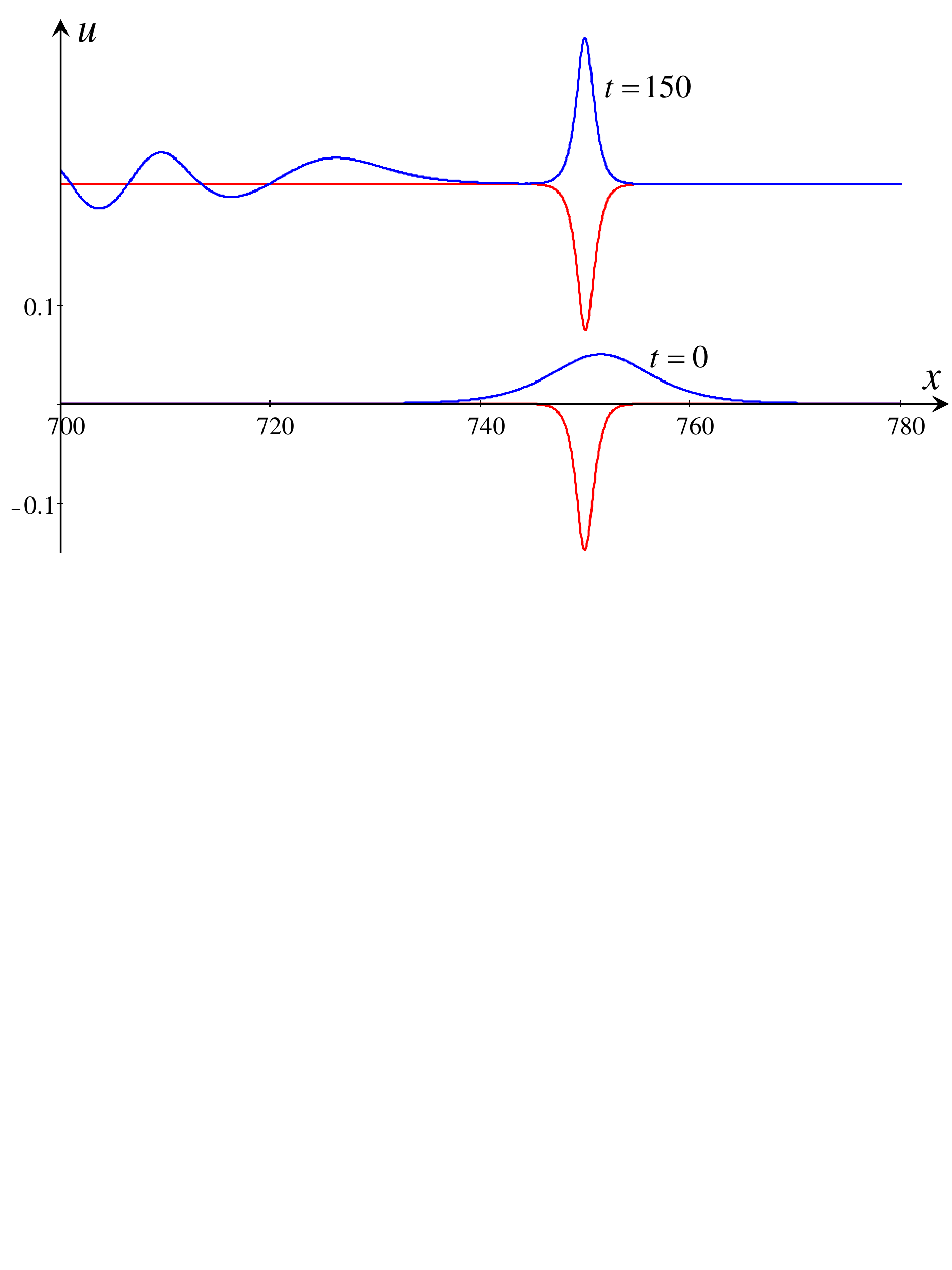}} %
	\vspace*{-6cm}
	\caption{Formation of a stationary soliton (blue line at $t = 150$) by the negative forcing (red lines) from a small-amplitude KdV soliton at $t = 0$. The initial soliton was slightly shifted to the right from the centre of forcing well.} %
	\label{Fig17}
\end{figure}

A similar situation occurs when, for the initial condition, a small-amplitude soliton is placed within the forcing well. The soliton quickly evolves into the statio\-nary soliton captured in the centre of forcing and emits a quasi-linear dispersive wave train (see Fig. \ref{Fig17}).

\begin{figure}[h!]
	\centerline{\includegraphics[width=8cm]{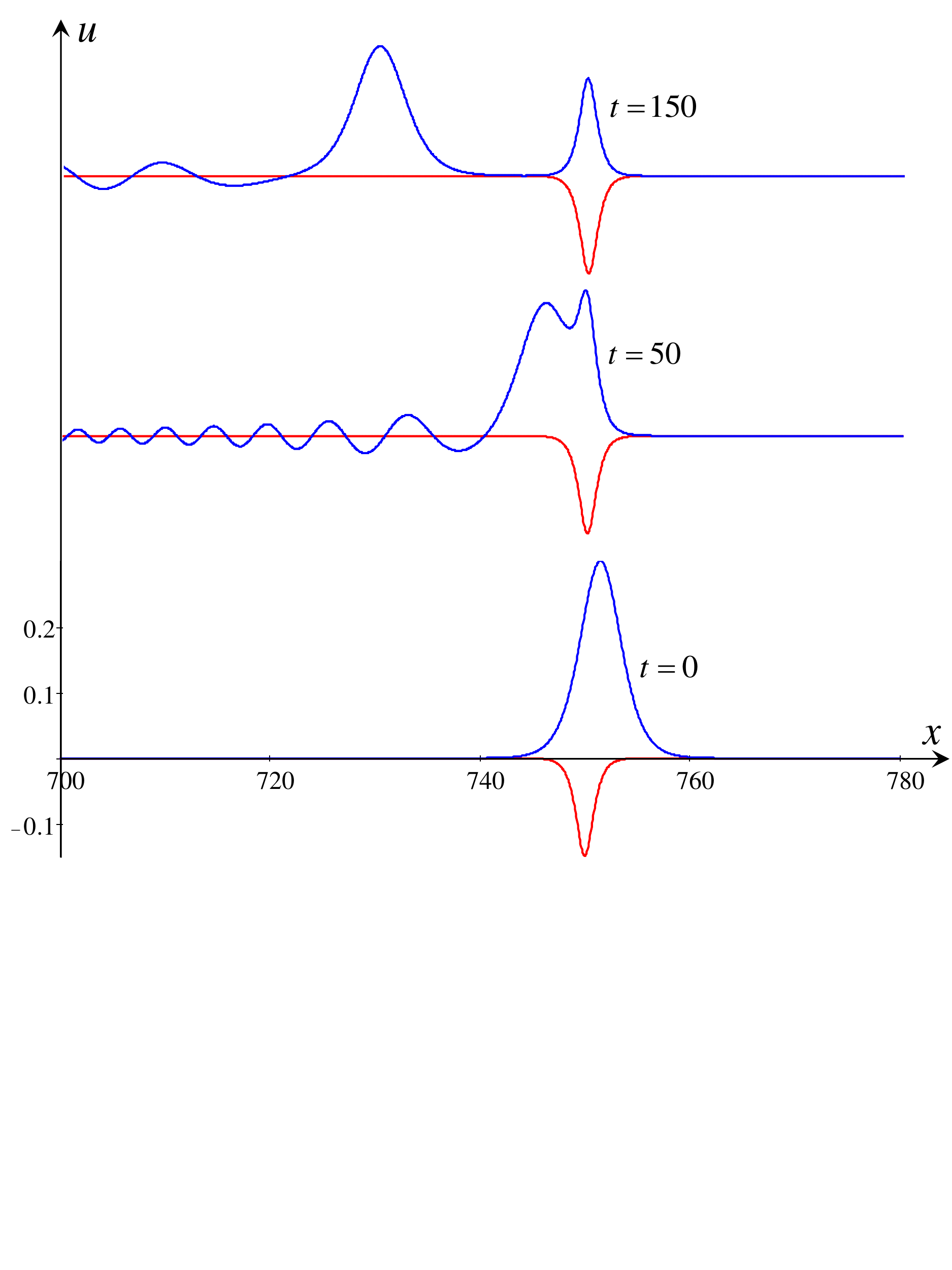}} %
	\vspace*{-3.5cm}
	\caption{Formation of the stationary soliton (blue line at $t = 150$) by the negative forcing (red lines) from the big-amplitude KdV soliton at $t = 0$. The initial soliton was slightly shifted to the right from the forcing centre.} %
	\label{Fig18}
\end{figure}

If, however, the amplitude of the initial soliton placed within the forcing zone is big, then the soliton splits under the action of forcing, so that one of its portions evolves into the stationary soliton captured in the centre of the forcing well, whereas another portion forms a soliton with different parameters freely moving with its own speed outside of the forcing zone. This process is accompanied by a quasi-linear dispersive wave train (see Fig. \ref{Fig18}). Such splitting and forming of a secondary soliton is beyond the range of applicability of the asymptotic theory.  
\begin{figure}[h!]
	\centerline{
	\includegraphics[width=11.cm]{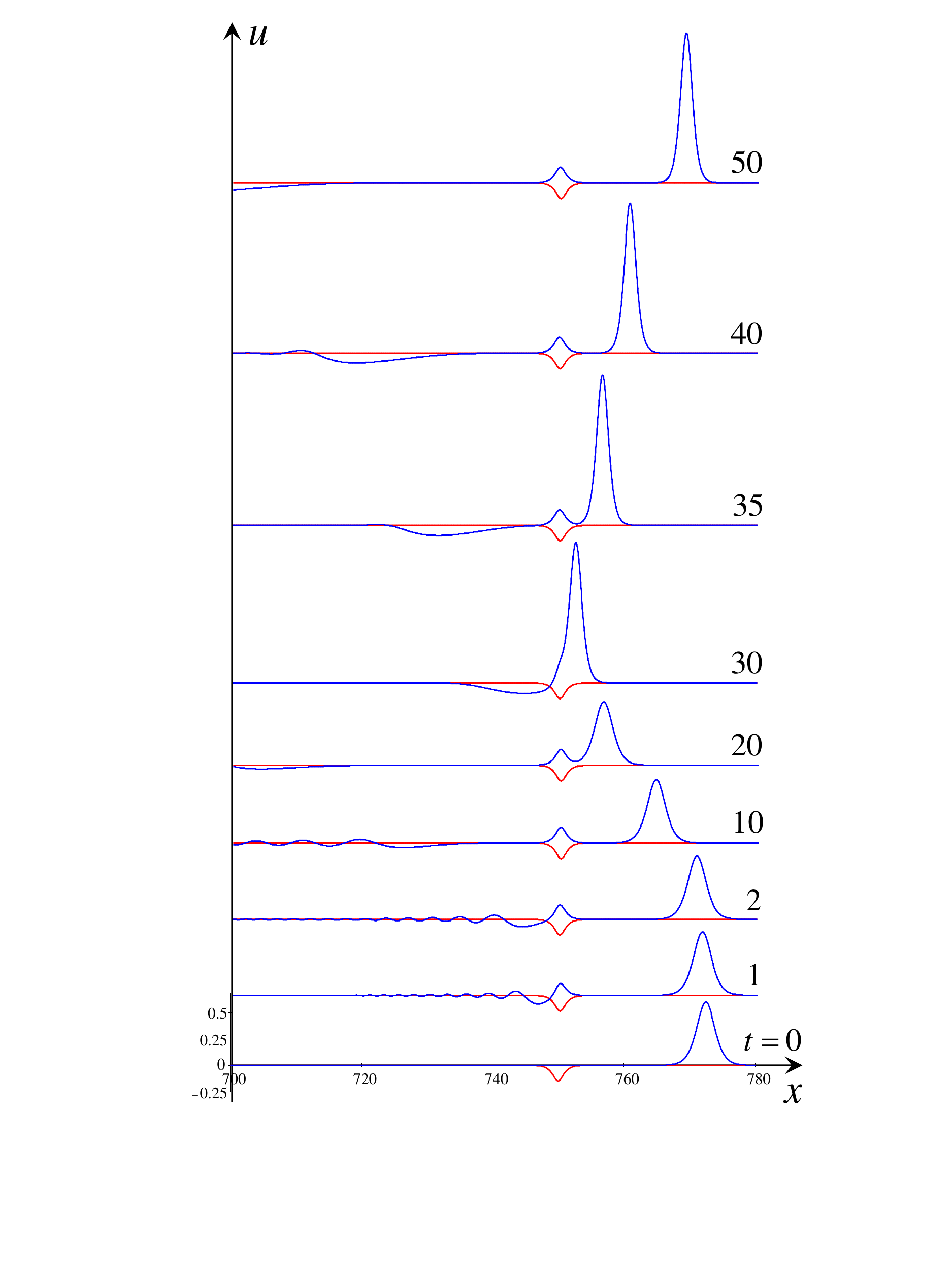}} %
	\vspace*{-2cm}
	\caption{Interaction of external KdV soliton approaching from the right with the negative forcing (red lines).} %
	\label{Fig19}
\end{figure}

In the case when a KdV soliton was placed initially outside of the forcing zone, we observed in numerical study both the reflection from the forcing and transition trough the forcing, as the asymptotic theory predicts for the moderate and big amplitude KdV solitons. Figure \ref{Fig19} illustrates the process of soliton reflection when it approaches the forcing from the right; this corresponds to the reflecting regime shown in Fig. \ref{ErmStep_Fig_13}d) on the right of the node. Because the forcing is attractive, it generates a stationary soliton from a noise, as was described above and shown in Fig. \ref{Fig16}. Therefore the external soliton shown in  Fig. \ref{Fig19} actually interacts with the forcing carrying a trapped stationary soliton. It is clearly seen in this figure that while the external soliton approaches the forcing, a small-amplitude trapped soliton forms by $t = 10$. Then the external soliton interacts with the forcing and soliton inside it and reflects back with a greater amplitude.

\begin{figure}[h!]
	\centerline{\hspace*{2cm}\includegraphics[width=11cm]{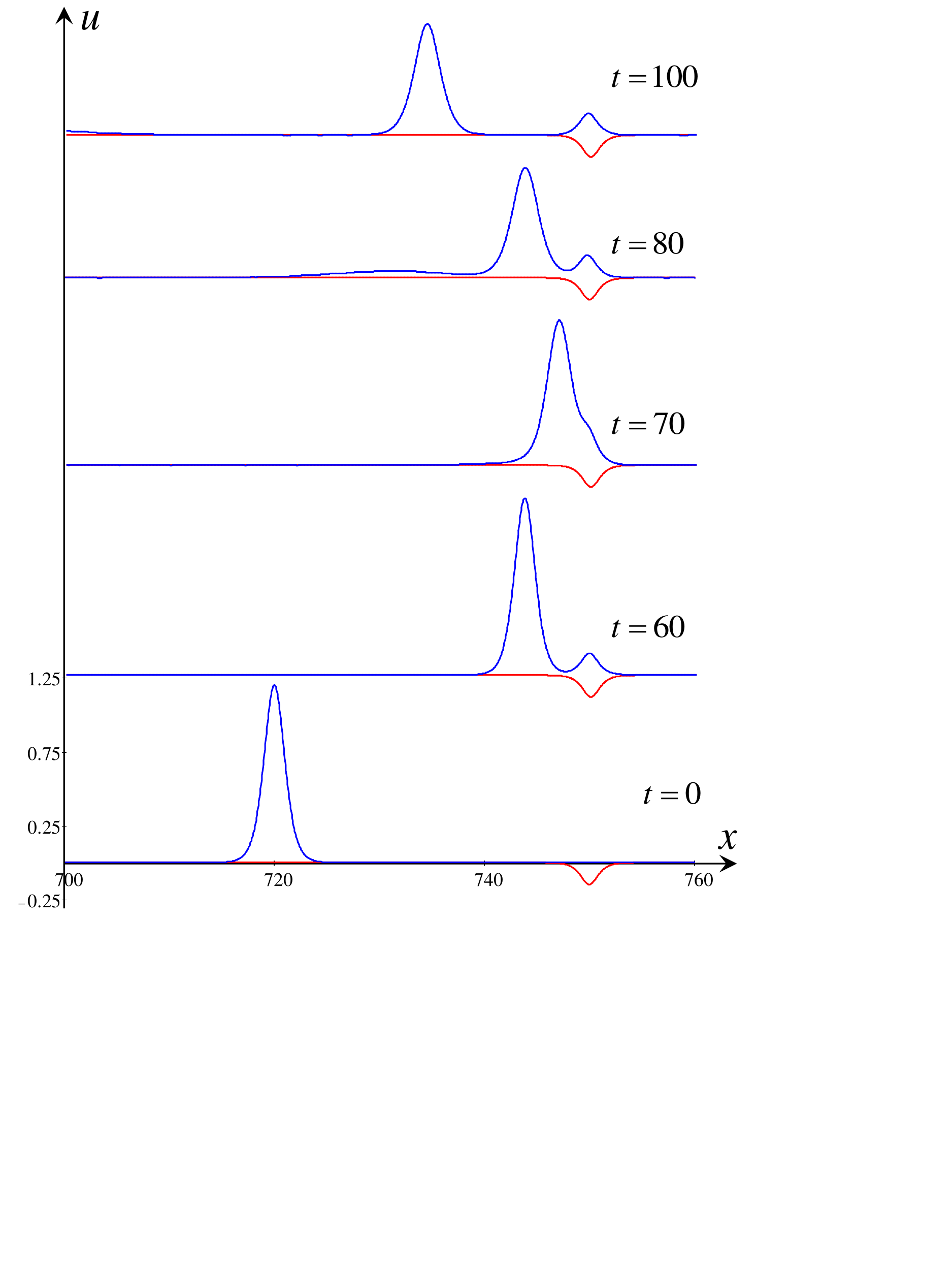}} %
	\vspace*{-4cm}
	\caption{Interaction of external KdV soliton approaching from the left with the negative forcing (red lines).} %
	\label{Fig20}
\end{figure}

A similar phenomenon occurs when a soliton approaches the forcing from the left as shown in Fig. \ref{Fig20}. In this figure one can see again that a stationary soliton emerges within the forcing from a noise while the external soliton approaches the forcing. Then the external soliton interacts with the forcing carrying the trapped stationary soliton and reflects back with a smaller amplitude emitting a small-amplitude wave in front of it. This corres\-ponds to the reflecting regime shown in the phase plane of Fig. \ref{ErmStep_Fig_13}d) -- see the phase trajectories on the left of the node.

\begin{figure}[h!]
	\centerline{\hspace*{2cm}\includegraphics[width=11cm]{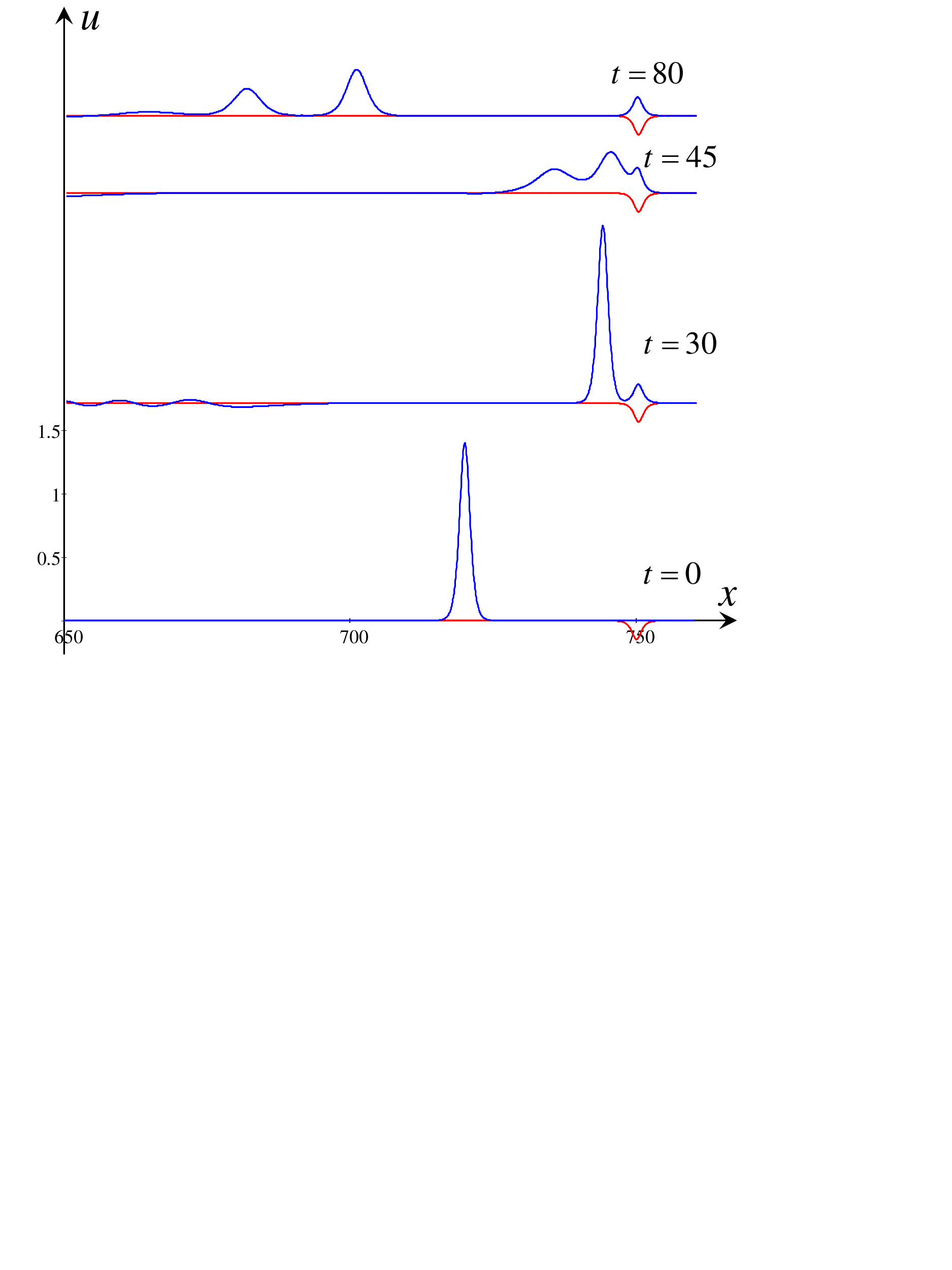}} %
	\vspace*{-7cm}
	\caption{Breakdown of an incident external KdV soliton into three solitons after reflection from the negative forcing (red lines).} %
	\label{Fig21}
\end{figure}

If the amplitude of external soliton is relatively big, then after reflection from the forcing it breaks into seve\-ral solitons as shown in Fig. \ref{Fig21}. The amplitudes of se\-condary solitons are noticeably less than the amplitude of the initial soliton; this agrees with the phase trajectories shown on the left from the node in the phase plane of Fig. \ref{ErmStep_Fig_13}d). The process of soliton breakdown onto secondary solitary waves after reflection from the forcing is not described by the asymptotic theory in its current form.

\begin{figure}[h!]
	\centerline{\hspace*{-0.3cm}\includegraphics[width=9cm]{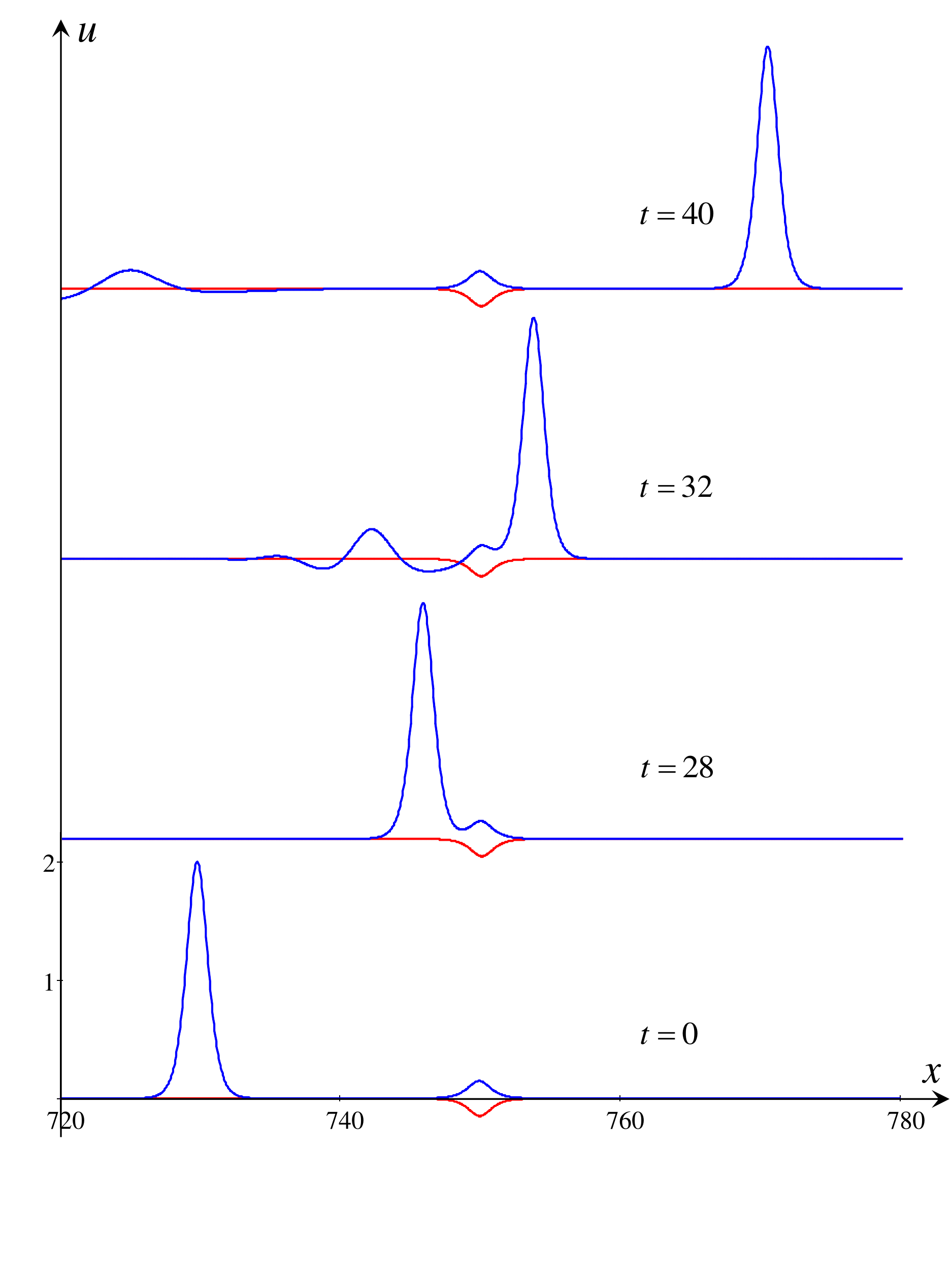}} %
	\vspace*{-1cm}
	\caption{Transition of incident external KdV soliton through the negative forcing (red lines).} %
	\label{Fig22}
\end{figure}

When the amplitude of external KdV soliton is too big, then the soliton simply passes through the forcing zone containing a stationary soliton and emits quasi-linear wave train. After that the soliton freely moves ahead as shown in Fig. \ref{Fig22}. This agrees with the transient phase trajectories shown above the node in the phase plane of Fig. \ref{ErmStep_Fig_13}d).

\section{Conclusion}
\label{Conclusion}

In this paper we have revised the asymptotic theory developed by Grimshaw and Pelinovsky with co-authors in the series of papers\cite{GPT1994, GPS1996, GPB1997, Pelin2000, GrimPel2002} to describe the dynamics of solitary waves in the KdV-like equations. In those papers only limiting cases were studied, either when the forcing is infinitely narrow in comparison with the initial KdV soliton and can be approximated by the Dirac $\delta$-function, or vice versa, when the initial KdV soliton is very narrow (approximated by the $\delta$-function) in comparison with the forcing of KdV-soliton shape. In our paper we consider an arbitrary relationship between the width of initial KdV soliton and external forcing. We present several examples of forced KdV equation which admit exact analytical solutions both stationary and non-stationary. 

In the case of small-amplitude forcing we have presented the asymptotic analysis based on equations derived in the papers cited above and have shown that in many cases solutions of approximate equations can be solved analytically, albeit the solutions look very cumbersome. In the limiting cases of very narrow or very wide forcing our results converge to those obtained in the papers by Grimshaw and Pelinovsky\cite{GPT1994}. In the meantime, we show that there are some physically interesting regimes which were missed in their papers due to approximations of soliton and forcing by the $\delta$-function. In particular, the equilibrium state of a stable focus in Fig. \ref{ErmStep_Fig_3}b) was mistakenly identified as a centre. Physically this implies that a soliton could oscillate with an arbitrary amplitude around the centre whereas in fact, the soliton quickly approaches a stable state moving synchronously with the forcing. Secondly, the repulsive regimes, when external solitary waves reflect from the forcing, were missed in that paper. Such regimes are clearly seen in the right lower corners of phase plane shown in Figs. \ref{ErmStep_Fig_3}a), \ref{ErmStep_Fig_10}b), and \ref{ErmStep_Fig_16}a),b), as well as illustrated in Figs. \ref{Fig19} and \ref{Fig20}. One of the most interesting regimes discovered in this paper is the permanent generation of solitary waves with a random amplitudes on a rear slope of a wide forcing as shown in Fig. \ref{Fig14}. This effect deserves further study which will undertaken in the nearest future.

The results obtained are important in view of their applications to physical phenomena occurring when external perturbations generate pressure fields capable of exciting and supporting solitary waves. This may happen, for example, when moving atmospheric pressure generates surface waves, or a slow-moving ship generates internal waves, or when atmospheric waves are generated behind high obstacles (for example, mountain ridges or other elevations). A similar phenomena can occur in the oceans when currents flow around underwater obstacles and generate surface and internal waves. The results obtained are applicable to other areas of physics, such as plasma physics and Bose--Einstein condensate, where the highly universal forced Korteweg--de Vries equation is used.

\section*{Acknowledgments}
A.E. acknowledges the financial support obtained from the Australian Government Research Training Program Scholarship. Y.S. acknowledges the funding of this study from the State task program in the sphere of scientific activity of the Ministry of Education and Science of the Russian Federation (Project No. 5.1246.2017/4.6) and grant of President of the Russian Federation for state support of leading Scientific Schools of the Russian Fe\-deration (NSH-2685.2018.5).

\section{Citations and References}%
\label{sec:endnotes}

\end{document}